\documentclass[aps,10pt,prl,superscriptaddress,twocolumn,footinbib,notitlepage,tightenlines,showpacs]{revtex4-1}

\usepackage{tensor}
\usepackage{amsmath}
\usepackage{amsthm}
\usepackage{graphicx}
\usepackage{float}

\usepackage{comment}
\usepackage{amsfonts}
\usepackage{bm}
\usepackage{braket}
\usepackage{xcolor}
\usepackage[colorlinks=true,linkcolor=cyan,urlcolor=teal,citecolor=cyan]{hyperref}

\usepackage{enumerate}
\usepackage{mathtools}
\usepackage{bbm}

\usepackage{quantikz2}
\usepackage{mathdots}

\newcommand{\ii}{\mathrm{i}}

\newcommand{\s}{S} 
\newcommand{\n}{N} 

\begin{document}
\title{Encrypted Qubits can be Cloned}

\makeatletter
\let\inserttitle\@title
\makeatother

\author{Koji Yamaguchi}
\altaffiliation{Current affiliation: Department of Informatics, Faculty of Information Science and Electrical Engineering,
Kyushu University, 744 Motooka, Nishi-ku, Fukuoka, 819-0395, Japan}
\email{koji.yamaguchi@inf.kyushu-u.ac.jp}

\affiliation{Department of Applied Mathematics, University of Waterloo, Waterloo, ON N2L 3G1, Canada}
\affiliation{Department of Communication Engineering and Informatics,
University of Electro-Communications, 1-5-1 Chofugaoka, Chofu, Tokyo, 182-8585, Japan}

\author{Achim Kempf}
\email{akempf@uwaterloo.ca}
\affiliation{Department of Applied Mathematics, University of Waterloo, Waterloo, ON N2L 3G1, Canada}
\affiliation{Department of Physics, University of Waterloo, Waterloo, ON N2L 3G1, Canada}
\affiliation{Institute for Quantum Computing, University of Waterloo, Waterloo, ON N2L 3G1, Canada}
\affiliation{Perimeter Institute for Theoretical Physics, Waterloo, Ontario N2L 2Y5, Canada}

\begin{abstract}
We show that \it encrypted cloning \rm
of unknown quantum states is possible. Any number of encrypted clones of
a qubit can be created through a 
unitary transformation, and each of the encrypted
clones can be decrypted through a unitary transformation. 
The decryption of an encrypted clone consumes
the decryption key, i.e., only one decryption is possible,
in agreement with the no-cloning theorem. 
Encrypted cloning represents a new paradigm that provides a form of redundancy, parallelism or scalability where direct duplication is forbidden by the no-cloning theorem. For example, a possible application of encrypted cloning 
is to enable encrypted quantum multi-cloud storage. 

\end{abstract}

\maketitle

The no-cloning theorem \cite{wootters_single_1982,dieks_communication_1982} shows that it is impossible to create an identical clone of an unknown quantum state, a fact that fundamentally limits the processing of quantum information. The theorem arises from the unitarity of quantum mechanics and shares a close relationship with the no-signaling principle \cite{herbert_flash-superluminal_1982,gisin_quantum_1998,navez_cloning_2003,masanes_general_2006}. The no-cloning theorem has spurred further related research, such as broadcasting of mixed states \cite{barnum_noncommuting_1996}, imperfect cloning \cite{buzek_quantum_1996,gisin_optimal_1997,brus_optimal_1998,werner_optimal_1998,keyl_optimal_1999}, and probabilistic cloning \cite{duan_probabilistic_1998,pati_quantum_1999} (see e.g., \cite{scarani_quantum_2005,fan_quantum_2014} for review). The implications of the no-cloning theorem are far-reaching, spanning a wide array of research areas in both quantum information science and fundamental physics, including quantum communication \cite{bennett_capacities_1997,giovannetti_broadband_2003,giovannetti_entanglement_2003,caruso_degradability_2006,holevo_entanglement-breaking_2008}, quantum error correction \cite{gottesman_stabilizer_1997}, quantum cryptography \cite{bennett_quantum_1984,gisin_quantum_1997,cerf_security_2002,gisin_quantum_2002,scarani_security_2009}, quantum estimation \cite{brus_optimal_1999,bae_asymptotic_2006}, and black hole physics \cite{hayden_black_2007,braunstein_better_2013}. 

In  this context, we here show that \it encrypted  cloning \rm of an unknown quantum state is possible. Crucially, this protocol generates multiple redundant, indirectly accessible encrypted copies, rather than multiple simultaneously accessible identical quantum states, thus strictly adhering to the no-cloning theorem.

To this end, we explicitly show that any number, $n>1$, of encrypted clones of a qubit, $A$, can be produced through a unitary transformation, and that each of the encrypted clones can be decrypted through a unitary transformation. The decryption of an encrypted clone consumes the decryption key, i.e., only one decryption is possible, in agreement with the no-cloning theorem. The number of gate operations needed for the encryption and decryption scales linearly with the number of clones.  

\begin{figure*}
    \centering
    \includegraphics[width=0.8\linewidth]{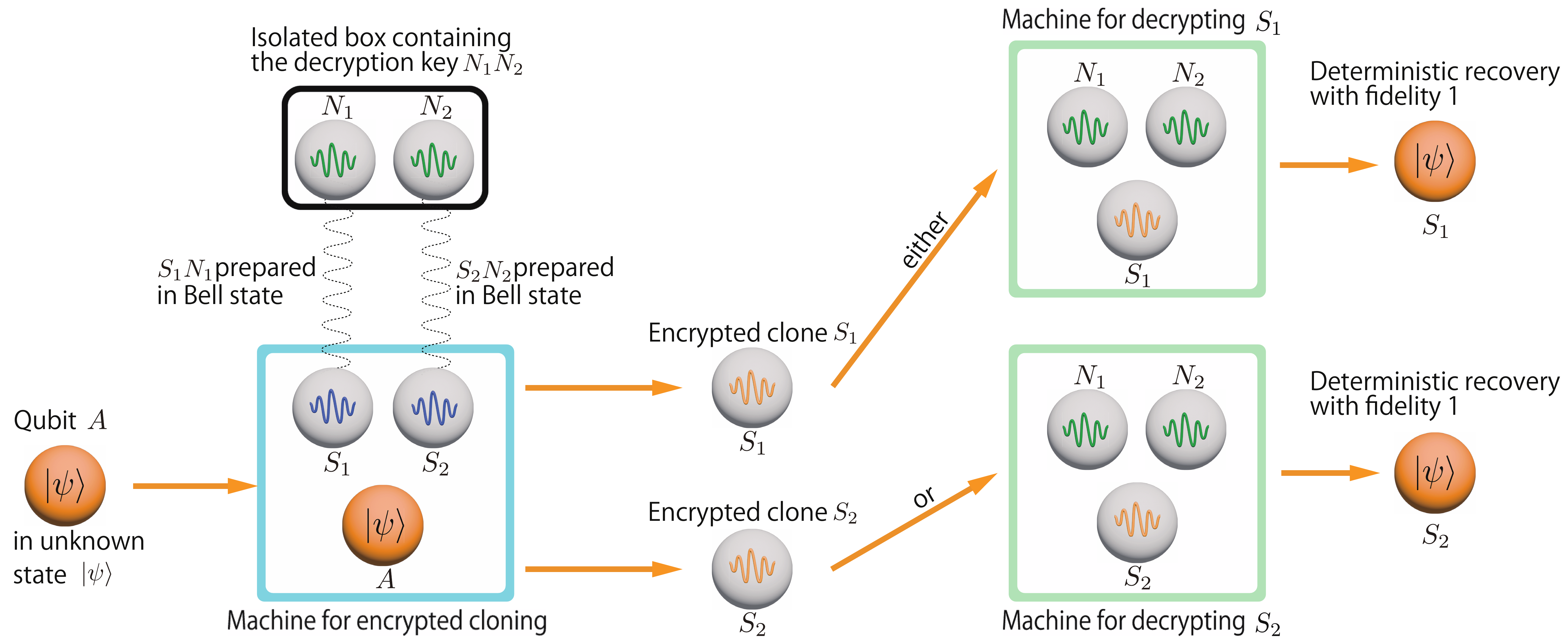}
    \caption{The protocol for $n=2$. Qubits whose reduced state is maximally mixed are represented by spheres displaying fluctuations. The initial maximal mixedness of $S_1$ and $S_2$, which stems from their being prepared in Bell states with $N_1$ and $N_2$ respectively, provides the quantum noise for the encryption. $N_1$ and $N_2$ keep a record of this quantum noise and can, therefore, later be used to de-noise or decrypt either $S_1$ or $S_2$. Crucially, the decryption machine consumes $N_1$ and $N_2$, so that only one decryption can be performed. Therefore, only one unencrypted version of the original state of $A$ can exist at a time, which enables consistency with the no-cloning theorem.}
    \label{fig:encrypted_cloning}
\end{figure*}

The new method of encrypted cloning (see Fig.~\ref{fig:encrypted_cloning}) begins with the preparation of $n$ pairs of maximally entangled qubits, $(\s_i,\n_i)$,  $i=1...n$, where we will refer to the $\s_i$ and $\n_i$ as signal qubits and noise qubits respectively. 
The method of encrypted cloning then has $A$ interact with all signal qubits $\s_i, ~i=1,...,n$ through a unitary operator $U^{(n)}_{\mathrm{enc}}$, that acts nontrivially only on qubit $A$ and the signal qubits. Through this encoding operation, complete information about qubit $A$ gets imprinted into each of the signal qubits $\s_i$. At the same time, each of these imprints is encrypted due to the quantum noise in the $\s_i$ from their initial maximal entanglement with the $\n_i$. We will call the imprinted signal qubits $\s_i$ the encrypted clones of the original state of $A$.  

The noise qubits $\n_i,~i=1...n$ do not take part in the process of encrypted cloning, i.e., $U^{(n)}_{\mathrm{enc}}$ acts as the identity on the Hilbert space of the $\n_i$. 
Therefore, the $\n_i$ acquire neither classical nor quantum information about $A$. 
Instead, the role of the noise qubits is to keep a record of the quantum noise in the signal qubits so that, together, the set of all noise qubits $\n_i,~i=1...n$ forms the encryption key. By using and thereby consuming this key, any one and only one of the encrypted clones $\s_i$ can then be decrypted to reproduce the original state of qubit $A$.    
The decryption or `denoising' of an encrypted clone $\s_k$ is accomplished by a unitary, $U^{(n)}_{\mathrm{dec}}$, that acts nontrivially only on $\s_k$ and the noise qubits $\n_i,~i=1...n$, and it reproduces the original state of qubit $A$ in qubit $\s_k$.

\noindent\textit{\textbf{Example application: Quantum Encrypted Multi-Cloud Storage.}}---

Let us consider a scenario in which the owner of quantum data requires for security reasons that these data are stored (a) off-site, (b) redundantly in multiple quantum clouds to protect, e.g., from hardware failures, and (c) encrypted, with the key kept by the owner on-site. 

The new method of encrypted cloning can be used for this purpose. To this end, the owner of a qubit, $A$, creates $n$ encrypted clones and distributes them to $n$ quantum clouds for storage. The owner can then discard the original qubit $A$. The owner retains the $n$ noise qubits, which together serve as the key. Condition a) is fulfilled because on-site are only the noise qubits and they do not carry information about qubit $A$. Condition b) is fulfilled since the owner can decrypt (i.e., denoise) any one of the redundantly stored encrypted clones to recover qubit $A$, by using the noise qubits. Condition c) is fulfilled since each of the encrypted clones that are hosted off-site in one of the quantum clouds is maximally mixed and independent of the initial state of $A$. 

\noindent\textit{\textbf{The method of encrypted cloning:}}--- 
We consider $n$  qubits $\s_i,~i=1...n$, called signal qubits, and $n$ qubits $\n_i,~i=1...n$, called noise qubits with each pair $(\s_i,\n_i)$ prepared in this maximally entangled state:
\begin{align}
    \ket{\phi}_{\s_i\n_i}\coloneqq \frac{1}{\sqrt{2}}\left(\ket{0}_{\s_i}\ket{0}_{\n_i}+\ket{1}_{\s_i}\ket{1}_{\n_i}\right)\label{eq:initial_state_Bs}.
\end{align}
We denote the identity operator and the Pauli operators for the $i$'th signal and noise qubits by $\{\sigma_\mu^{(\s_i)}\}_{\mu=0,1,2,3}$ and $\{\sigma_\mu^{(\n_i)}\}_{\mu=0,1,2,3}$ respectively, in their $\{\ket{0},\ket{1}\}$ basis.

The owner of quantum data, say Alice, possesses a qubit $A$ and aims to create $n$ encrypted clones of the state of qubit $A$. To this end, Alice lets her qubit $A$ interact with the $n$ signal qubits $\{\s_i\}_{i=1}^n$ through the following unitary encrypted cloning operation:
\begin{align}
    U_{\mathrm{enc}}^{(n)}\coloneqq e^{-\frac{\pi\ii }{4} \sigma_{1}^{(A)}\otimes \left(\bigotimes_{i=1}^n\sigma_{1}^{(\s_i)}\right)}e^{-\frac{\pi\ii}{4} \sigma_{3}^{(A)}\otimes \left(\bigotimes_{i=1}^n\sigma_{3}^{(\s_i)}\right)}. 
    \label{eq:encoding_optimal}
\end{align}
Here, the superscript of each Pauli operator denotes the subsystem that this Pauli operator acts on. The unitary $U^{(n)}_{\mathrm{enc}}$ acts as the identity operator on the noise qubits' tensor factor of the Hilbert space. Therefore, when $U^{(n)}_{\mathrm{enc}}$ is compiled into a succession of native gates of a quantum processor, no gates act on the noise qubits, i.e., the noise qubits are noninteracting ancillas that may as well be kept physically apart. Hence, the noise qubits carry neither classical nor quantum information about Alice's qubit. 

\noindent\textit{\textbf{Decryption: Proof of full recovery.}}---
After the process of encrypted cloning, Bob (which can also be Alice) attempts to recover the initial state of qubit $A$ from a subset of the signal qubits and noise qubits. Correspondingly, we consider this quantum channel from Alice to Bob:
\begin{align}
    &\mathcal{N}_{A\to B}^{(n)}(\rho_A)\nonumber\\
    &\coloneqq \mathrm{Tr}_{\overline{B}}\left(U_{\mathrm{enc}}^{(n)}\left(\rho_A\otimes\left(\bigotimes_{i=1}^n\phi_{\s_i\n_i}\right)\right)U_{\mathrm{enc}}^{(n)\dag} \right).\label{eq:encoding_channel}
\end{align}
Here, we defined $\phi_{\s_i\n_i}\coloneqq\ket{\phi}\bra{\phi}_{\s_i\n_i}$. Also, we let $B$ denote the qubit(s) that Bob chooses to use to try to reproduce the original state $\rho_A$ of qubit $A$. We let $\overline{B}$ denote the complementary system to $B$. For example, if $B=\s_1$, then $\overline{B}=A\s_2\cdots \s_n\n_1\cdots \n_n$.  
Our aim is to show that if Bob uses any one signal qubit, $\s_i$, and all noise qubits, $\n_1,...,\n_n$, then Bob can retrieve Alice's qubit perfectly, in the sense that the channel from Alice to Bob is of full quantum capacity $C_Q$:
\begin{align}
    C_Q(\mathcal{N}_{A\to \s_i\n_1\n_2\cdots \n_n}^{(n)})=1\label{eq:channel_capacity_copy}.
\end{align}
We will see that the capacity drops to zero if Bob omits even only one noise qubit.

We prove Eq.~\eqref{eq:channel_capacity_copy} by explicitly constructing a decrypting operation (we provide an alternative proof using coherent information in the Supplementary Material ~\cite{sm}). To this end, we show that Alice's initial state $\rho_A$ is fully recovered from $\s_1\n_1\n_2\cdots \n_n$ by the decrypting operation $\mathcal{D}_{\s_1\n_1\n_2\cdots \n_n\to \s_1}^{(n)}$ defined for $n>1$ by
\begin{align}
    &\mathcal{D}_{\s_1\n_1\n_2\cdots \n_n\to \s_1}^{(n)}(\cdots)\coloneqq \mathrm{Tr}_{\n_1\n_2\cdots \n_n}\left(U_{\mathrm{dec}}^{(n)}(\cdots )U_{\mathrm{dec}}^{(n)\dag}\right),\nonumber\\
&U_{\mathrm{dec}}^{(n)}\coloneqq \sum_{\mu=0}^3\alpha_\mu\left(\ket{\phi_\mu}\bra{\phi_\mu}_{\s_1\n_1}\right)\otimes \left(\bigotimes_{j=2}^n\sigma_{\mu}^{(\n_j)\top}\right) \label{eq:decoding_op}. 
\end{align}
Here, $\ket{\phi_\mu}_{\s_i\n_i}\coloneqq \sigma_{\mu}^{(\s_i)}\otimes \mathbb{I}^{(\n_i)}\ket{\phi}_{\s_i\n_i}$, $\top$ denotes the transpose operation with respect to the computational basis $\{\ket{0},\ket{1}\}$, and the coefficients are defined by $\alpha_0\coloneqq 1$, $\alpha_1=\alpha_3\coloneqq\ii$ and $\alpha_2\coloneqq -\ii^{n+1}$. 

First, it is straightforward to check that $U_{\mathrm{dec}}^{(n)}$ is unitary by using the fact that $\{\ket{\phi_\mu}_{\s_i\n_i}\}_{\mu=0}^3$ forms an orthonormal basis of a two-qubit system and $|\alpha_\mu|^2=1$. Therefore, $\mathcal{D}_{\s_1\n_1\n_2\cdots \n_n\to \s_1}^{(n)}$ is a quantum channel. Second, from Eq.~\eqref{eq:encoding_optimal}, we find
\begin{align}
    &U_{\mathrm{enc}}^{(n)}=\left(\cos(\pi/4)\mathbb{I}-\ii\sin(\pi/4) \sigma_{1}^{(A)}\otimes \left(\bigotimes_{i=1}^n\sigma_{1}^{(\s_i)}\right)\right) \nonumber\\
    &\times \left(\cos(\pi/4)\mathbb{I}-\ii\sin(\pi/4) \sigma_{3}^{(A)}\otimes \left(\bigotimes_{i=1}^n\sigma_{3}^{(\s_i)}\right)\right)\nonumber\\
    &=\frac{1}{2}\sum_{\mu=0}^3\alpha_\mu^{-1}\sigma_{\mu}^{(A)}\otimes \left(\bigotimes_{i=1}^n\sigma_{\mu}^{(\s_i)}\right),
\end{align}
where we have used $\sigma_1^{\otimes (n+1)}\sigma_3^{\otimes (n+1)}=(-\ii)^{n+1}\sigma_2^{\otimes (n+1)}$. 
Therefore, we get
\begin{align}
    &U_{\mathrm{enc}}^{(n)}\left(\ket{\psi}_A\otimes\left(\bigotimes_{i=1}^n\ket{\phi}_{\s_i\n_i}\right)\right)\nonumber\\
    &=\frac{1}{2}\sum_{\mu=0}^3\alpha_\mu^{-1} \sigma_\mu^{(A)}\ket{\psi}_A\otimes\left(\bigotimes_{i=1}^n\ket{\phi_\mu}_{\s_i\n_i}\right)\label{eq:state_after_encoding},
\end{align}
implying that
\begin{align}
    &\left(U_{\mathrm{dec}}^{(n)}\otimes \mathbb{I}_{A\s_2\s_3\cdots \s_n}\right)U_{\mathrm{enc}}^{(n)}\left(\ket{\psi}_A\otimes\left(\bigotimes_{i=1}^n\ket{\phi}_{\s_i\n_i}\right)\right)\nonumber\\
    &=\frac{1}{2}\left(\sum_{\mu=0}^3\sigma_{\mu}^{(A)}\otimes \sigma_{\mu}^{(\s_1)}\otimes \mathbb{I}\right)\ket{\psi}_A\otimes \bigotimes_{i=1}^n\left(\ket{\phi}_{\s_i\n_i}\right),\label{eq:state_after_decoding_total}
\end{align}
where we have used $\sigma_\mu^{(\s_i)}\otimes \sigma_{\mu}^{(\n_i)\top}\ket{\phi}_{\s_i\n_i}=\ket{\phi}_{\s_i\n_i}$.
Since $\frac{1}{2}\sum_{\mu=0}^3\sigma_{\mu}^{(A)}\otimes \sigma_{\mu}^{(\s_1)}$ is a SWAP operation between $A$ and $\s_1$, and since $\mathcal{N}_{A\to \s_1\n_1\n_2\cdots \n_n}^{(n)}(\cdots)=\mathrm{Tr}_{A\s_2\s_3\cdots \s_n}(U_{\mathrm{enc}}^{(n)}(\cdots\bigotimes_{i=1}^n\phi_{\s_i\n_i})U_{\mathrm{enc}}^{(n)\dag})$, we conclude that
\begin{align}
    \mathcal{D}_{\s_1\n_1\n_2\cdots \n_n\to \s_1}^{(n)}\circ \mathcal{N}_{A\to \s_1\n_1\n_2\cdots \n_n}^{(n)}(\ket{\psi}\bra{\psi}_{A})=\ket{\psi}\bra{\psi}_{\s_1}\label{eq:decryption_oneshot}
\end{align}
holds for any initial pure state $\ket{\psi}_A$ of Alice's qubit. Therefore, Alice's initial state can be retrieved perfectly from $\s_1\n_1\n_2\cdots \n_n$. As one can also decrypt the same information from $\s_i\n_1\n_2\cdots \n_n$ due to the symmetry under exchanging the roles of $\s_1\n_1$ and $\s_i\n_i$ in the decoding operation, Eq.~\eqref{eq:channel_capacity_copy} is proven for any $i$. 
Therefore, Bob can recover the original state of qubit $A$ using one arbitrary signal qubit and all noise qubits. One signal qubit alone would not be sufficient to recover the original state of $A$ since  
otherwise it could also be recovered from the remaining signal qubits due to symmetry, thereby violating the no-cloning theorem. Indeed,
\begin{align}
    C_Q(\mathcal{N}_{A\to \s_i}^{(n)})=0 \,\,\,\, \text{ (no-cloning)}\label{eq:channel_capacity_no_cloning}
\end{align}
since $\mathcal{N}_{A\to \s_i}^{(n)}$ is anti-degradable due to the permutation symmetry among signal qubits, and the quantum capacity vanishes for any anti-degradable channel \cite{bennett_capacities_1997,giovannetti_broadband_2003,giovannetti_entanglement_2003,caruso_degradability_2006,holevo_entanglement-breaking_2008}. We remark that the present results were inspired by our prior work \cite{yamaguchi_superadditivity_2020,ahmadzadegan_neural_2021} on classical and quantum denoising.

In conclusion, each signal qubit contains an encrypted but perfect clone of Alice's original qubit, $A$. As we show in the Supplementary Material~\cite{sm}, each clone is perfectly encrypted in the sense that it is noisy to the extent that each individually is in a maximally mixed state. Yet, each encrypted clone is also a perfect copy of the original state of $A$ in the sense that the original state of $A$ can be perfectly recovered from the encrypted clone by `denoising' or `decoding' or `decrypting' - we will here use these three terms interchangeably - using only the noise qubits which never interacted with $A$ and therefore contain no classical or quantum information about $A$. 

Let us clarify that consistency with the no-cloning theorem demands that, after decrypting one signal qubit, it is impossible to reuse the noise qubits to decrypt another signal qubit. Indeed, after decoding, as Eq.~\eqref{eq:state_after_decoding_total} shows, the state of the $(n-1)$ unused signal qubits and $n$ noise qubits is independent of Alice's information. Of course, it would not violate no-cloning if, after denoising one signal qubit, Bob runs the unitary decoding operation Eq.~\eqref{eq:decoding_op} in reverse and then chooses to denoise another signal qubit. 

We remark that, trivially, the pair of encoding and decoding unitary operations, as in Eqs.~\eqref{eq:encoding_optimal} and \eqref{eq:decoding_op}, can be replaced by rotated versions. For example, $\sigma_{3}^{(A)}\otimes \left(\bigotimes_{i=1}^n\sigma_{3}^{(\s_i)}\right)$ can be replaced with $\sigma_{2}^{(A)}\otimes \left(\bigotimes_{i=1}^n\sigma_{2}^{(\s_i)}\right)$ while in the encoding operation in Eq.~\eqref{eq:encoding_optimal}, the coefficients $\alpha_\mu$ in Eq.~\eqref{eq:state_after_encoding} change into $\alpha_0= 1$, $\alpha_1=\alpha_2=\ii$ and $\alpha_3= -(-\ii)^{n+1}$. Then, Alice's quantum information can be retrieved using the decoding operation~\eqref{eq:decoding_op} with these coefficients. 

More generally, the encoding and decoding unitaries of Eqs.~\eqref{eq:encoding_optimal} and \eqref{eq:decoding_op} can be replaced by any of the infinitely many unitaries that act the same way on the initial states and on the encoded states respectively. 

The choice of encoding and decoding unitaries can be important in practical applications, where the encoding and decoding unitaries need to be decomposed, i.e., compiled into the universal native one- and two-qubit gates of a given quantum processor \cite{divincenzo_two-bit_1995,lloyd_almost_1995,barenco_elementary_1995}. As we also show in the Supplementary Material~\cite{sm}, the number of two-qubit gates that is required to implement the copying and retrieval operations scales well, namely by growing only linearly with $n$. 

We here remark that in recent experimental work in progress with Christian Tutschku, Leon Rullk\"otter, Sean Wagner, and Ibrahim Shehzad, 
we successfully implemented encrypted cloning and the subsequent decryption for up to $n=10$ on a quantum processor \cite{implemented}. 

For variants of the protocol, see Appendix A in End Matter.

\noindent\textit{\textbf{Relationship to other phenomena.}}---\newline
\textbf{1. From classical one-time pads to quantum multi-time pads} \par

The encryption aspect of the new encrypted cloning method can also be understood in analogy to classical one-time pad (OTP) cryptography. This is because the quantum information that is imprinted on the signal qubits is perfectly masked by noise, as if from an OTP, the noise being here the quantum noise from the initial maximal entanglement between the signal and noise qubits. The noise qubits, therefore, represent an analog of a one-time-pad, except that in the quantum case, the pad can actually be re-used without compromising security. This is because the encrypted cloning method encrypts and decrypts without measurements, unitarily, in such a way that, after decryption, the $n$ maximally entangled pairs are restored and are, therefore, available again for encryption. 

\textbf{2. Consistency with the no quantum summoning theorem}\par

Quantum summoning \cite{kent_quantum_2012,kent_no-summoning_2013,hayden_summoning_2016,adlam_quantum_2016} is an adversarial game played in spacetime. At an event $X$, Player 1 gives  Player 2 a qubit $A$. Then, at some event $Y_i$ from a pre-agreed upon set of events in the causal future of $X$, Player 1 asks Player 2 to produce qubit $A$. Player 2 is able to win this challenge if all events $Y_i$ are timelike to another, because then Player 2 can transport or teleport qubit $A$ from one $Y_i$ to the next until it is summoned by Player 1. The no quantum summoning theorem states that Player 2 cannot possess an always-winning strategy if some of the $Y_i$ are spacelike to another, as this would contradict no-cloning. 

While encrypted cloning is clearly consistent with the no quantum summoning theorem, the application of encrypted cloning to quantum encrypted multi-cloud storage also shows that a variant of quantum summoning is possible also for spacelike separated $Y_i$: Player 2 deposits encrypted clones at each of the $Y_i$. Then, Player 1 can summon qubit $A$ at any $Y_i$ of her choosing - even if some or all of the $Y_i$ are spacelike separated from another - provided that Player 1 brings the decryption key, i.e., the noise qubits. Recall that the noise qubits do not carry any classical or quantum information about $A$. 

\textbf{3. Consistency with quantum secret sharing}\par
The new encrypted cloning method is consistent not only with the no-cloning theorem, but also with quantum secret sharing \cite{cleve_how_1999,gottesman_theory_2000}. To see this, we need to consider what in quantum secret sharing is called the access structure, namely the list of so-called authorized or unauthorized sets of parties (subsystems) from which a secret qubit can or cannot be recovered respectively. In encrypted cloning, any subsystem containing one pair of signal and noise qubits and one half of each of the remaining $(n-1)$ pairs is authorized, while any subsystem of $(n-1)$ pairs or $n$ noise qubits is unauthorized. As is easy to check, the two necessary and sufficient conditions required for quantum secret sharing \cite{gottesman_theory_2000}, namely (a) that the complement of any authorized set is unauthorized and (b) monotonicity, i.e., that any superset of an authorized set is authorized, are readily verified.  

Since any quantum secret sharing scheme can correct erasure errors on the complementary system to an authorized set \cite{cleve_how_1999}, encrypted cloning can, therefore, also be viewed as an error-correcting code. For example, after encrypted cloning, the original state of qubit $A$ can be retrieved even if $(n-1)$ signal qubits are lost. 

\textbf{4. Consistency with entanglement monogamy}\par
Let us assume that qubit $A$ is initially maximally entangled with an ancilla qubit $\tilde{A}$. After the encrypted cloning, $\tilde{A}$ is, therefore, simultaneously maximally entangled with each set of qubits from which the original state of $A$ can be recovered. 

This fact does not violate entanglement monogamy \cite{coffman_distributed_2000} since the monogamy argument only applies to disjoint sets of subsystems while in encrypted cloning every set of qubits from which the original state of $A$ can be reconstructed overlaps with any other such set. 

\textbf{5. Relation to the Hayden-Preskill model} \par
Similar to encrypted cloning, which assumes that initially we have a number of maximally entangled pairs of signal and noise qubits, in the case of black holes, the degrees of freedom of the black hole and the degrees of freedom of the Hawking radiation that it has emitted by the Page time can be assumed to be maximally entangled. Then, similar to the imprinting of a qubit $A$ into the signal qubits in encrypted cloning, when sending a qubit $A$ into a black hole, it may become imprinted into the degrees of freedom in the black hole through fast scrambling. In encrypted cloning, $A$ can be recovered from even just a single one of the encrypted clones $S_i$. Similarly in spirit, in the Hayden-Preskill model, \cite{hayden_black_2007}, as soon as some more of the degrees of freedom (perhaps analogous to the imprinted signal qubits) evaporate from the black hole, qubit $A$ can be recovered with the help of all of the prior degrees of freedom of the Hawking radiation (analogous to the noise qubits).  

\noindent\textit{\textbf{Outlook.}}---
While technical tools used in encrypted cloning, such as the manipulation of multipartite entanglement, are also foundational in quantum error correction (QEC) \cite{knill_theory_1997,roffe_quantum_2019,brun_quantum_2020,georgescu_25_2020} and quantum secret sharing (QSS) \cite{cleve_how_1999,gottesman_theory_2000}, the operational goals and applications differ. Encrypted cloning is not designed to protect against computational errors (as in QEC) or to restrict information access among agents (as in QSS). Instead, it enables the creation of encrypted yet decryptable clones of an unknown quantum state — a new functionality, relevant in contexts where classical copying is essential, but forbidden quantum mechanically by the no-cloning theorem. (And the encrypted cloning protocol achieves this entirely unitarily, without syndrome measurements or classical communication, and permits arbitrary selection of the decrypted clone.)

For example, as we discussed, the new encrypted cloning paradigm naturally enables quantum multi-cloud storage and it will be very interesting to explore its extension to quantum multi-cloud parallel homomorphic, or `blind', computation. \rm

More generally, the paradigm of encrypted cloning can also be seen as a method to evade a constraint imposed by unitarity, namely in this case the no-cloning constraint. To this end, encrypted cloning enlarges the system with ancillas, thereby leaving the original system an open system. Quantum noise is then introduced through maximum entanglement with ancillas - but in such a way that later denoising is possible. 

This is reminiscent of how unitarity also forces quantum linear amplifiers to introduce quantum noise \cite{haus_quantum_1962,caves_quantum_1982}. It will be interesting, therefore, to explore whether an encrypted cloning type approach can be used to develop new quantum amplifier architectures in which the necessary quantum noise is introduced through maximally entangled pairs - in such a way that halves of these pairs may later serve to at least partially denoise a quantum amplified signal. Similarly, also the no-programming theorem \cite{nielsen1997programmable,yang2020optimal} arises from unitarity and an encrypted cloning approach may, therefore, provide a new perspective. 

It should also be very interesting to explore an encrypted cloning approach for quantum sensing. Useful for sensing could also be the fact that, for large $n$, the ability to recover qubit $A$ can be extremely sensitive to any interactions that a to-be-probed system may have with the $n$ noise qubits.

Vice versa, in encrypted cloning we have enhanced robustness in the transmission of the signal qubits, in the sense that if $n$ signal qubits are sent, it suffices that even just one arrives to achieve full recovery. For example, one may envision a quantum radar type setup, see \cite{lloyd_enhanced_2008}, in which $n$ noise qubits in photons are kept as idlers while $n$ signal photons are emitted, of which only one signal photon needs to be received.    

Finally, it will also be interesting to generalize encrypted cloning from qubits to qudits, and to explore the corresponding scaling of the minimal resources needed.


\let\oldaddcontentsline\addcontentsline
\renewcommand{\addcontentsline}[3]{}

\begin{acknowledgments}
K.Y. acknowledges support from the JSPS Overseas Research Fellowships and JSPS Grants-in-Aid for Scientific Research No.JP24KJ0085. 
A.K. acknowledges support through the Dieter Schwarz Foundation, a Discovery Grant of the National Science and Engineering Council of Canada (NSERC), an Applied Quantum Computing Challenge Grant from the National Research Council of Canada (NRC), a Discovery Project grant of the Australian Research Council (ARC) and support from Perimeter Institute which is supported in part by the Government of Canada through the Department of Innovation, Science and Industry Canada and by the Province of Ontario through the Ministry of Colleges and Universities. 
\end{acknowledgments}


\begin{thebibliography}{63}%
\makeatletter
\providecommand \@ifxundefined [1]{%
 \@ifx{#1\undefined}
}%
\providecommand \@ifnum [1]{%
 \ifnum #1\expandafter \@firstoftwo
 \else \expandafter \@secondoftwo
 \fi
}%
\providecommand \@ifx [1]{%
 \ifx #1\expandafter \@firstoftwo
 \else \expandafter \@secondoftwo
 \fi
}%
\providecommand \natexlab [1]{#1}%
\providecommand \enquote  [1]{``#1''}%
\providecommand \bibnamefont  [1]{#1}%
\providecommand \bibfnamefont [1]{#1}%
\providecommand \citenamefont [1]{#1}%
\providecommand \href@noop [0]{\@secondoftwo}%
\providecommand \href [0]{\begingroup \@sanitize@url \@href}%
\providecommand \@href[1]{\@@startlink{#1}\@@href}%
\providecommand \@@href[1]{\endgroup#1\@@endlink}%
\providecommand \@sanitize@url [0]{\catcode `\\12\catcode `\$12\catcode
  `\&12\catcode `\#12\catcode `\^12\catcode `\_12\catcode `\%12\relax}%
\providecommand \@@startlink[1]{}%
\providecommand \@@endlink[0]{}%
\providecommand \url  [0]{\begingroup\@sanitize@url \@url }%
\providecommand \@url [1]{\endgroup\@href {#1}{\urlprefix }}%
\providecommand \urlprefix  [0]{URL }%
\providecommand \Eprint [0]{\href }%
\providecommand \doibase [0]{http://dx.doi.org/}%
\providecommand \selectlanguage [0]{\@gobble}%
\providecommand \bibinfo  [0]{\@secondoftwo}%
\providecommand \bibfield  [0]{\@secondoftwo}%
\providecommand \translation [1]{[#1]}%
\providecommand \BibitemOpen [0]{}%
\providecommand \bibitemStop [0]{}%
\providecommand \bibitemNoStop [0]{.\EOS\space}%
\providecommand \EOS [0]{\spacefactor3000\relax}%
\providecommand \BibitemShut  [1]{\csname bibitem#1\endcsname}%
\let\auto@bib@innerbib\@empty
\bibitem [{\citenamefont {Wootters}\ and\ \citenamefont
  {Zurek}(1982)}]{wootters_single_1982}%
  \BibitemOpen
  \bibfield  {author} {\bibinfo {author} {\bibfnamefont {W.~K.}\ \bibnamefont
  {Wootters}}\ and\ \bibinfo {author} {\bibfnamefont {W.~H.}\ \bibnamefont
  {Zurek}},\ }\href {\doibase 10.1038/299802a0} {\bibfield  {journal} {\bibinfo
   {journal} {Nature}\ }\textbf {\bibinfo {volume} {299}},\ \bibinfo {pages}
  {802} (\bibinfo {year} {1982})}\BibitemShut {NoStop}%
\bibitem [{\citenamefont {Dieks}(1982)}]{dieks_communication_1982}%
  \BibitemOpen
  \bibfield  {author} {\bibinfo {author} {\bibfnamefont {D.}~\bibnamefont
  {Dieks}},\ }\href {\doibase 10.1016/0375-9601(82)90084-6} {\bibfield
  {journal} {\bibinfo  {journal} {Physics Letters A}\ }\textbf {\bibinfo
  {volume} {92}},\ \bibinfo {pages} {271} (\bibinfo {year} {1982})}\BibitemShut
  {NoStop}%
\bibitem [{\citenamefont {Herbert}(1982)}]{herbert_flash-superluminal_1982}%
  \BibitemOpen
  \bibfield  {author} {\bibinfo {author} {\bibfnamefont {N.}~\bibnamefont
  {Herbert}},\ }\href {\doibase 10.1007/BF00729622} {\bibfield  {journal}
  {\bibinfo  {journal} {Foundations of Physics}\ }\textbf {\bibinfo {volume}
  {12}},\ \bibinfo {pages} {1171} (\bibinfo {year} {1982})}\BibitemShut
  {NoStop}%
\bibitem [{\citenamefont {Gisin}(1998)}]{gisin_quantum_1998}%
  \BibitemOpen
  \bibfield  {author} {\bibinfo {author} {\bibfnamefont {N.}~\bibnamefont
  {Gisin}},\ }\href {\doibase 10.1016/S0375-9601(98)00170-4} {\bibfield
  {journal} {\bibinfo  {journal} {Physics Letters A}\ }\textbf {\bibinfo
  {volume} {242}},\ \bibinfo {pages} {1} (\bibinfo {year} {1998})}\BibitemShut
  {NoStop}%
\bibitem [{\citenamefont {Navez}\ and\ \citenamefont
  {Cerf}(2003)}]{navez_cloning_2003}%
  \BibitemOpen
  \bibfield  {author} {\bibinfo {author} {\bibfnamefont {P.}~\bibnamefont
  {Navez}}\ and\ \bibinfo {author} {\bibfnamefont {N.~J.}\ \bibnamefont
  {Cerf}},\ }\href {\doibase 10.1103/PhysRevA.68.032313} {\bibfield  {journal}
  {\bibinfo  {journal} {Physical Review A}\ }\textbf {\bibinfo {volume} {68}},\
  \bibinfo {pages} {032313} (\bibinfo {year} {2003})}\BibitemShut {NoStop}%
\bibitem [{\citenamefont {Masanes}\ \emph {et~al.}(2006)\citenamefont
  {Masanes}, \citenamefont {Acin},\ and\ \citenamefont
  {Gisin}}]{masanes_general_2006}%
  \BibitemOpen
  \bibfield  {author} {\bibinfo {author} {\bibfnamefont {L.}~\bibnamefont
  {Masanes}}, \bibinfo {author} {\bibfnamefont {A.}~\bibnamefont {Acin}}, \
  and\ \bibinfo {author} {\bibfnamefont {N.}~\bibnamefont {Gisin}},\ }\href
  {\doibase 10.1103/PhysRevA.73.012112} {\bibfield  {journal} {\bibinfo
  {journal} {Physical Review A}\ }\textbf {\bibinfo {volume} {73}},\ \bibinfo
  {pages} {012112} (\bibinfo {year} {2006})}\BibitemShut {NoStop}%
\bibitem [{\citenamefont {Barnum}\ \emph {et~al.}(1996)\citenamefont {Barnum},
  \citenamefont {Caves}, \citenamefont {Fuchs}, \citenamefont {Jozsa},\ and\
  \citenamefont {Schumacher}}]{barnum_noncommuting_1996}%
  \BibitemOpen
  \bibfield  {author} {\bibinfo {author} {\bibfnamefont {H.}~\bibnamefont
  {Barnum}}, \bibinfo {author} {\bibfnamefont {C.~M.}\ \bibnamefont {Caves}},
  \bibinfo {author} {\bibfnamefont {C.~A.}\ \bibnamefont {Fuchs}}, \bibinfo
  {author} {\bibfnamefont {R.}~\bibnamefont {Jozsa}}, \ and\ \bibinfo {author}
  {\bibfnamefont {B.}~\bibnamefont {Schumacher}},\ }\href {\doibase
  10.1103/PhysRevLett.76.2818} {\bibfield  {journal} {\bibinfo  {journal}
  {Physical Review Letters}\ }\textbf {\bibinfo {volume} {76}},\ \bibinfo
  {pages} {2818} (\bibinfo {year} {1996})}\BibitemShut {NoStop}%
\bibitem [{\citenamefont {Bužek}\ and\ \citenamefont
  {Hillery}(1996)}]{buzek_quantum_1996}%
  \BibitemOpen
  \bibfield  {author} {\bibinfo {author} {\bibfnamefont {V.}~\bibnamefont
  {Bužek}}\ and\ \bibinfo {author} {\bibfnamefont {M.}~\bibnamefont
  {Hillery}},\ }\href {\doibase 10.1103/PhysRevA.54.1844} {\bibfield  {journal}
  {\bibinfo  {journal} {Physical Review A}\ }\textbf {\bibinfo {volume} {54}},\
  \bibinfo {pages} {1844} (\bibinfo {year} {1996})}\BibitemShut {NoStop}%
\bibitem [{\citenamefont {Gisin}\ and\ \citenamefont
  {Massar}(1997)}]{gisin_optimal_1997}%
  \BibitemOpen
  \bibfield  {author} {\bibinfo {author} {\bibfnamefont {N.}~\bibnamefont
  {Gisin}}\ and\ \bibinfo {author} {\bibfnamefont {S.}~\bibnamefont {Massar}},\
  }\href {\doibase 10.1103/PhysRevLett.79.2153} {\bibfield  {journal} {\bibinfo
   {journal} {Physical Review Letters}\ }\textbf {\bibinfo {volume} {79}},\
  \bibinfo {pages} {2153} (\bibinfo {year} {1997})}\BibitemShut {NoStop}%
\bibitem [{\citenamefont {Bruß}\ \emph {et~al.}(1998)\citenamefont {Bruß},
  \citenamefont {DiVincenzo}, \citenamefont {Ekert}, \citenamefont {Fuchs},
  \citenamefont {Macchiavello},\ and\ \citenamefont
  {Smolin}}]{brus_optimal_1998}%
  \BibitemOpen
  \bibfield  {author} {\bibinfo {author} {\bibfnamefont {D.}~\bibnamefont
  {Bruß}}, \bibinfo {author} {\bibfnamefont {D.~P.}\ \bibnamefont
  {DiVincenzo}}, \bibinfo {author} {\bibfnamefont {A.}~\bibnamefont {Ekert}},
  \bibinfo {author} {\bibfnamefont {C.~A.}\ \bibnamefont {Fuchs}}, \bibinfo
  {author} {\bibfnamefont {C.}~\bibnamefont {Macchiavello}}, \ and\ \bibinfo
  {author} {\bibfnamefont {J.~A.}\ \bibnamefont {Smolin}},\ }\href {\doibase
  10.1103/PhysRevA.57.2368} {\bibfield  {journal} {\bibinfo  {journal}
  {Physical Review A}\ }\textbf {\bibinfo {volume} {57}},\ \bibinfo {pages}
  {2368} (\bibinfo {year} {1998})}\BibitemShut {NoStop}%
\bibitem [{\citenamefont {Werner}(1998)}]{werner_optimal_1998}%
  \BibitemOpen
  \bibfield  {author} {\bibinfo {author} {\bibfnamefont {R.~F.}\ \bibnamefont
  {Werner}},\ }\href {\doibase 10.1103/PhysRevA.58.1827} {\bibfield  {journal}
  {\bibinfo  {journal} {Physical Review A}\ }\textbf {\bibinfo {volume} {58}},\
  \bibinfo {pages} {1827} (\bibinfo {year} {1998})}\BibitemShut {NoStop}%
\bibitem [{\citenamefont {Keyl}\ and\ \citenamefont
  {Werner}(1999)}]{keyl_optimal_1999}%
  \BibitemOpen
  \bibfield  {author} {\bibinfo {author} {\bibfnamefont {M.}~\bibnamefont
  {Keyl}}\ and\ \bibinfo {author} {\bibfnamefont {R.~F.}\ \bibnamefont
  {Werner}},\ }\href {\doibase 10.1063/1.532887} {\bibfield  {journal}
  {\bibinfo  {journal} {Journal of Mathematical Physics}\ }\textbf {\bibinfo
  {volume} {40}},\ \bibinfo {pages} {3283} (\bibinfo {year}
  {1999})}\BibitemShut {NoStop}%
\bibitem [{\citenamefont {Duan}\ and\ \citenamefont
  {Guo}(1998)}]{duan_probabilistic_1998}%
  \BibitemOpen
  \bibfield  {author} {\bibinfo {author} {\bibfnamefont {L.-M.}\ \bibnamefont
  {Duan}}\ and\ \bibinfo {author} {\bibfnamefont {G.-C.}\ \bibnamefont {Guo}},\
  }\href {\doibase 10.1103/PhysRevLett.80.4999} {\bibfield  {journal} {\bibinfo
   {journal} {Physical Review Letters}\ }\textbf {\bibinfo {volume} {80}},\
  \bibinfo {pages} {4999} (\bibinfo {year} {1998})}\BibitemShut {NoStop}%
\bibitem [{\citenamefont {Pati}(1999)}]{pati_quantum_1999}%
  \BibitemOpen
  \bibfield  {author} {\bibinfo {author} {\bibfnamefont {A.~K.}\ \bibnamefont
  {Pati}},\ }\href {\doibase 10.1103/PhysRevLett.83.2849} {\bibfield  {journal}
  {\bibinfo  {journal} {Physical Review Letters}\ }\textbf {\bibinfo {volume}
  {83}},\ \bibinfo {pages} {2849} (\bibinfo {year} {1999})}\BibitemShut
  {NoStop}%
\bibitem [{\citenamefont {Scarani}\ \emph {et~al.}(2005)\citenamefont
  {Scarani}, \citenamefont {Iblisdir}, \citenamefont {Gisin},\ and\
  \citenamefont {Acín}}]{scarani_quantum_2005}%
  \BibitemOpen
  \bibfield  {author} {\bibinfo {author} {\bibfnamefont {V.}~\bibnamefont
  {Scarani}}, \bibinfo {author} {\bibfnamefont {S.}~\bibnamefont {Iblisdir}},
  \bibinfo {author} {\bibfnamefont {N.}~\bibnamefont {Gisin}}, \ and\ \bibinfo
  {author} {\bibfnamefont {A.}~\bibnamefont {Acín}},\ }\href {\doibase
  10.1103/RevModPhys.77.1225} {\bibfield  {journal} {\bibinfo  {journal}
  {Reviews of Modern Physics}\ }\textbf {\bibinfo {volume} {77}},\ \bibinfo
  {pages} {1225} (\bibinfo {year} {2005})}\BibitemShut {NoStop}%
\bibitem [{\citenamefont {Fan}\ \emph {et~al.}(2014)\citenamefont {Fan},
  \citenamefont {Wang}, \citenamefont {Jing}, \citenamefont {Yue},
  \citenamefont {Shi}, \citenamefont {Zhang},\ and\ \citenamefont
  {Mu}}]{fan_quantum_2014}%
  \BibitemOpen
  \bibfield  {author} {\bibinfo {author} {\bibfnamefont {H.}~\bibnamefont
  {Fan}}, \bibinfo {author} {\bibfnamefont {Y.-N.}\ \bibnamefont {Wang}},
  \bibinfo {author} {\bibfnamefont {L.}~\bibnamefont {Jing}}, \bibinfo {author}
  {\bibfnamefont {J.-D.}\ \bibnamefont {Yue}}, \bibinfo {author} {\bibfnamefont
  {H.-D.}\ \bibnamefont {Shi}}, \bibinfo {author} {\bibfnamefont {Y.-L.}\
  \bibnamefont {Zhang}}, \ and\ \bibinfo {author} {\bibfnamefont {L.-Z.}\
  \bibnamefont {Mu}},\ }\href {\doibase 10.1016/j.physrep.2014.06.004}
  {\bibfield  {journal} {\bibinfo  {journal} {Physics Reports}\ }\bibinfo
  {series} {Quantum cloning machines and the applications},\ \textbf {\bibinfo
  {volume} {544}},\ \bibinfo {pages} {241} (\bibinfo {year}
  {2014})}\BibitemShut {NoStop}%
\bibitem [{\citenamefont {Bennett}\ \emph {et~al.}(1997)\citenamefont
  {Bennett}, \citenamefont {DiVincenzo},\ and\ \citenamefont
  {Smolin}}]{bennett_capacities_1997}%
  \BibitemOpen
  \bibfield  {author} {\bibinfo {author} {\bibfnamefont {C.~H.}\ \bibnamefont
  {Bennett}}, \bibinfo {author} {\bibfnamefont {D.~P.}\ \bibnamefont
  {DiVincenzo}}, \ and\ \bibinfo {author} {\bibfnamefont {J.~A.}\ \bibnamefont
  {Smolin}},\ }\href {\doibase 10.1103/PhysRevLett.78.3217} {\bibfield
  {journal} {\bibinfo  {journal} {Physical Review Letters}\ }\textbf {\bibinfo
  {volume} {78}},\ \bibinfo {pages} {3217} (\bibinfo {year}
  {1997})}\BibitemShut {NoStop}%
\bibitem [{\citenamefont {Giovannetti}\ \emph
  {et~al.}(2003{\natexlab{a}})\citenamefont {Giovannetti}, \citenamefont
  {Lloyd}, \citenamefont {Maccone},\ and\ \citenamefont
  {Shor}}]{giovannetti_broadband_2003}%
  \BibitemOpen
  \bibfield  {author} {\bibinfo {author} {\bibfnamefont {V.}~\bibnamefont
  {Giovannetti}}, \bibinfo {author} {\bibfnamefont {S.}~\bibnamefont {Lloyd}},
  \bibinfo {author} {\bibfnamefont {L.}~\bibnamefont {Maccone}}, \ and\
  \bibinfo {author} {\bibfnamefont {P.~W.}\ \bibnamefont {Shor}},\ }\href
  {\doibase 10.1103/PhysRevA.68.062323} {\bibfield  {journal} {\bibinfo
  {journal} {Physical Review A}\ }\textbf {\bibinfo {volume} {68}},\ \bibinfo
  {pages} {062323} (\bibinfo {year} {2003}{\natexlab{a}})}\BibitemShut
  {NoStop}%
\bibitem [{\citenamefont {Giovannetti}\ \emph
  {et~al.}(2003{\natexlab{b}})\citenamefont {Giovannetti}, \citenamefont
  {Lloyd}, \citenamefont {Maccone},\ and\ \citenamefont
  {Shor}}]{giovannetti_entanglement_2003}%
  \BibitemOpen
  \bibfield  {author} {\bibinfo {author} {\bibfnamefont {V.}~\bibnamefont
  {Giovannetti}}, \bibinfo {author} {\bibfnamefont {S.}~\bibnamefont {Lloyd}},
  \bibinfo {author} {\bibfnamefont {L.}~\bibnamefont {Maccone}}, \ and\
  \bibinfo {author} {\bibfnamefont {P.~W.}\ \bibnamefont {Shor}},\ }\href
  {\doibase 10.1103/PhysRevLett.91.047901} {\bibfield  {journal} {\bibinfo
  {journal} {Physical Review Letters}\ }\textbf {\bibinfo {volume} {91}},\
  \bibinfo {pages} {047901} (\bibinfo {year} {2003}{\natexlab{b}})}\BibitemShut
  {NoStop}%
\bibitem [{\citenamefont {Caruso}\ and\ \citenamefont
  {Giovannetti}(2006)}]{caruso_degradability_2006}%
  \BibitemOpen
  \bibfield  {author} {\bibinfo {author} {\bibfnamefont {F.}~\bibnamefont
  {Caruso}}\ and\ \bibinfo {author} {\bibfnamefont {V.}~\bibnamefont
  {Giovannetti}},\ }\href {\doibase 10.1103/PhysRevA.74.062307} {\bibfield
  {journal} {\bibinfo  {journal} {Physical Review A}\ }\textbf {\bibinfo
  {volume} {74}},\ \bibinfo {pages} {062307} (\bibinfo {year}
  {2006})}\BibitemShut {NoStop}%
\bibitem [{\citenamefont {Holevo}(2008)}]{holevo_entanglement-breaking_2008}%
  \BibitemOpen
  \bibfield  {author} {\bibinfo {author} {\bibfnamefont {A.~S.}\ \bibnamefont
  {Holevo}},\ }\href {\doibase 10.1134/S0032946008030010} {\bibfield  {journal}
  {\bibinfo  {journal} {Problems of Information Transmission}\ }\textbf
  {\bibinfo {volume} {44}},\ \bibinfo {pages} {171} (\bibinfo {year}
  {2008})}\BibitemShut {NoStop}%
\bibitem [{\citenamefont {Gottesman}(1997)}]{gottesman_stabilizer_1997}%
  \BibitemOpen
  \bibfield  {author} {\bibinfo {author} {\bibfnamefont {D.~E.}\ \bibnamefont
  {Gottesman}},\ }\emph {\bibinfo {title} {Stabilizer {Codes} and {Quantum}
  {Error} {Correction}}},\ \href {\doibase 10.7907/rzr7-dt72} {\bibinfo {type}
  {{PhD} {Thesis}}},\ \bibinfo  {school} {California Institute of Technology}
  (\bibinfo {year} {1997})\BibitemShut {NoStop}%
\bibitem [{\citenamefont {Bennett}\ and\ \citenamefont
  {Brassard}(1984)}]{bennett_quantum_1984}%
  \BibitemOpen
  \bibfield  {author} {\bibinfo {author} {\bibfnamefont {C.~H.}\ \bibnamefont
  {Bennett}}\ and\ \bibinfo {author} {\bibfnamefont {G.}~\bibnamefont
  {Brassard}},\ }\href {\doibase 10.1016/j.tcs.2014.05.025} {\bibfield
  {journal} {\bibinfo  {journal} {Proceedings of the International Conference
  on Computers, Systems \& Signal Processing}\ ,\ \bibinfo {pages} {175}}
  (\bibinfo {year} {1984})}\BibitemShut {NoStop}%
\bibitem [{\citenamefont {Gisin}\ and\ \citenamefont
  {Huttner}(1997)}]{gisin_quantum_1997}%
  \BibitemOpen
  \bibfield  {author} {\bibinfo {author} {\bibfnamefont {N.}~\bibnamefont
  {Gisin}}\ and\ \bibinfo {author} {\bibfnamefont {B.}~\bibnamefont
  {Huttner}},\ }\href {\doibase 10.1016/S0375-9601(97)00083-2} {\bibfield
  {journal} {\bibinfo  {journal} {Physics Letters A}\ }\textbf {\bibinfo
  {volume} {228}},\ \bibinfo {pages} {13} (\bibinfo {year} {1997})}\BibitemShut
  {NoStop}%
\bibitem [{\citenamefont {Cerf}\ \emph {et~al.}(2002)\citenamefont {Cerf},
  \citenamefont {Bourennane}, \citenamefont {Karlsson},\ and\ \citenamefont
  {Gisin}}]{cerf_security_2002}%
  \BibitemOpen
  \bibfield  {author} {\bibinfo {author} {\bibfnamefont {N.~J.}\ \bibnamefont
  {Cerf}}, \bibinfo {author} {\bibfnamefont {M.}~\bibnamefont {Bourennane}},
  \bibinfo {author} {\bibfnamefont {A.}~\bibnamefont {Karlsson}}, \ and\
  \bibinfo {author} {\bibfnamefont {N.}~\bibnamefont {Gisin}},\ }\href
  {\doibase 10.1103/PhysRevLett.88.127902} {\bibfield  {journal} {\bibinfo
  {journal} {Physical Review Letters}\ }\textbf {\bibinfo {volume} {88}},\
  \bibinfo {pages} {127902} (\bibinfo {year} {2002})}\BibitemShut {NoStop}%
\bibitem [{\citenamefont {Gisin}\ \emph {et~al.}(2002)\citenamefont {Gisin},
  \citenamefont {Ribordy}, \citenamefont {Tittel},\ and\ \citenamefont
  {Zbinden}}]{gisin_quantum_2002}%
  \BibitemOpen
  \bibfield  {author} {\bibinfo {author} {\bibfnamefont {N.}~\bibnamefont
  {Gisin}}, \bibinfo {author} {\bibfnamefont {G.}~\bibnamefont {Ribordy}},
  \bibinfo {author} {\bibfnamefont {W.}~\bibnamefont {Tittel}}, \ and\ \bibinfo
  {author} {\bibfnamefont {H.}~\bibnamefont {Zbinden}},\ }\href {\doibase
  10.1103/RevModPhys.74.145} {\bibfield  {journal} {\bibinfo  {journal}
  {Reviews of Modern Physics}\ }\textbf {\bibinfo {volume} {74}},\ \bibinfo
  {pages} {145} (\bibinfo {year} {2002})}\BibitemShut {NoStop}%
\bibitem [{\citenamefont {Scarani}\ \emph {et~al.}(2009)\citenamefont
  {Scarani}, \citenamefont {Bechmann-Pasquinucci}, \citenamefont {Cerf},
  \citenamefont {Dušek}, \citenamefont {Lütkenhaus},\ and\ \citenamefont
  {Peev}}]{scarani_security_2009}%
  \BibitemOpen
  \bibfield  {author} {\bibinfo {author} {\bibfnamefont {V.}~\bibnamefont
  {Scarani}}, \bibinfo {author} {\bibfnamefont {H.}~\bibnamefont
  {Bechmann-Pasquinucci}}, \bibinfo {author} {\bibfnamefont {N.~J.}\
  \bibnamefont {Cerf}}, \bibinfo {author} {\bibfnamefont {M.}~\bibnamefont
  {Dušek}}, \bibinfo {author} {\bibfnamefont {N.}~\bibnamefont {Lütkenhaus}},
  \ and\ \bibinfo {author} {\bibfnamefont {M.}~\bibnamefont {Peev}},\ }\href
  {\doibase 10.1103/RevModPhys.81.1301} {\bibfield  {journal} {\bibinfo
  {journal} {Reviews of Modern Physics}\ }\textbf {\bibinfo {volume} {81}},\
  \bibinfo {pages} {1301} (\bibinfo {year} {2009})}\BibitemShut {NoStop}%
\bibitem [{\citenamefont {Bruß}\ and\ \citenamefont
  {Macchiavello}(1999)}]{brus_optimal_1999}%
  \BibitemOpen
  \bibfield  {author} {\bibinfo {author} {\bibfnamefont {D.}~\bibnamefont
  {Bruß}}\ and\ \bibinfo {author} {\bibfnamefont {C.}~\bibnamefont
  {Macchiavello}},\ }\href {\doibase 10.1016/S0375-9601(99)00099-7} {\bibfield
  {journal} {\bibinfo  {journal} {Physics Letters A}\ }\textbf {\bibinfo
  {volume} {253}},\ \bibinfo {pages} {249} (\bibinfo {year}
  {1999})}\BibitemShut {NoStop}%
\bibitem [{\citenamefont {Bae}\ and\ \citenamefont
  {Acín}(2006)}]{bae_asymptotic_2006}%
  \BibitemOpen
  \bibfield  {author} {\bibinfo {author} {\bibfnamefont {J.}~\bibnamefont
  {Bae}}\ and\ \bibinfo {author} {\bibfnamefont {A.}~\bibnamefont {Acín}},\
  }\href {\doibase 10.1103/PhysRevLett.97.030402} {\bibfield  {journal}
  {\bibinfo  {journal} {Physical Review Letters}\ }\textbf {\bibinfo {volume}
  {97}},\ \bibinfo {pages} {030402} (\bibinfo {year} {2006})}\BibitemShut
  {NoStop}%
\bibitem [{\citenamefont {Hayden}\ and\ \citenamefont
  {Preskill}(2007)}]{hayden_black_2007}%
  \BibitemOpen
  \bibfield  {author} {\bibinfo {author} {\bibfnamefont {P.}~\bibnamefont
  {Hayden}}\ and\ \bibinfo {author} {\bibfnamefont {J.}~\bibnamefont
  {Preskill}},\ }\href {\doibase 10.1088/1126-6708/2007/09/120} {\bibfield
  {journal} {\bibinfo  {journal} {Journal of High Energy Physics}\ }\textbf
  {\bibinfo {volume} {2007}},\ \bibinfo {pages} {120} (\bibinfo {year}
  {2007})}\BibitemShut {NoStop}%
\bibitem [{\citenamefont {Braunstein}\ \emph {et~al.}(2013)\citenamefont
  {Braunstein}, \citenamefont {Pirandola},\ and\ \citenamefont
  {Życzkowski}}]{braunstein_better_2013}%
  \BibitemOpen
  \bibfield  {author} {\bibinfo {author} {\bibfnamefont {S.~L.}\ \bibnamefont
  {Braunstein}}, \bibinfo {author} {\bibfnamefont {S.}~\bibnamefont
  {Pirandola}}, \ and\ \bibinfo {author} {\bibfnamefont {K.}~\bibnamefont
  {Życzkowski}},\ }\href {\doibase 10.1103/PhysRevLett.110.101301} {\bibfield
  {journal} {\bibinfo  {journal} {Physical Review Letters}\ }\textbf {\bibinfo
  {volume} {110}},\ \bibinfo {pages} {101301} (\bibinfo {year}
  {2013})}\BibitemShut {NoStop}%
\bibitem [{sm()}]{sm}%
  \BibitemOpen
  \href@noop {} {}\bibinfo {note} {See Supplemental Material for
  further details, which includes Refs. [55-63].}\BibitemShut {Stop}%
\bibitem [{\citenamefont {Yamaguchi}\ \emph {et~al.}(2020)\citenamefont
  {Yamaguchi}, \citenamefont {Ahmadzadegan}, \citenamefont {Simidzija},
  \citenamefont {Kempf},\ and\ \citenamefont
  {Martín-Martínez}}]{yamaguchi_superadditivity_2020}%
  \BibitemOpen
  \bibfield  {author} {\bibinfo {author} {\bibfnamefont {K.}~\bibnamefont
  {Yamaguchi}}, \bibinfo {author} {\bibfnamefont {A.}~\bibnamefont
  {Ahmadzadegan}}, \bibinfo {author} {\bibfnamefont {P.}~\bibnamefont
  {Simidzija}}, \bibinfo {author} {\bibfnamefont {A.}~\bibnamefont {Kempf}}, \
  and\ \bibinfo {author} {\bibfnamefont {E.}~\bibnamefont
  {Martín-Martínez}},\ }\href {\doibase 10.1103/PhysRevD.101.105009}
  {\bibfield  {journal} {\bibinfo  {journal} {Physical Review D}\ }\textbf
  {\bibinfo {volume} {101}},\ \bibinfo {pages} {105009} (\bibinfo {year}
  {2020})}\BibitemShut {NoStop}%
\bibitem [{\citenamefont {Ahmadzadegan}\ \emph {et~al.}(2021)\citenamefont
  {Ahmadzadegan}, \citenamefont {Simidzija}, \citenamefont {Li},\ and\
  \citenamefont {Kempf}}]{ahmadzadegan_neural_2021}%
  \BibitemOpen
  \bibfield  {author} {\bibinfo {author} {\bibfnamefont {A.}~\bibnamefont
  {Ahmadzadegan}}, \bibinfo {author} {\bibfnamefont {P.}~\bibnamefont
  {Simidzija}}, \bibinfo {author} {\bibfnamefont {M.}~\bibnamefont {Li}}, \
  and\ \bibinfo {author} {\bibfnamefont {A.}~\bibnamefont {Kempf}},\ }\href
  {\doibase 10.1038/s41598-021-00502-4} {\bibfield  {journal} {\bibinfo
  {journal} {Scientific Reports}\ }\textbf {\bibinfo {volume} {11}},\ \bibinfo
  {pages} {21624} (\bibinfo {year} {2021})}\BibitemShut {NoStop}%
\bibitem [{\citenamefont {DiVincenzo}(1995)}]{divincenzo_two-bit_1995}%
  \BibitemOpen
  \bibfield  {author} {\bibinfo {author} {\bibfnamefont {D.~P.}\ \bibnamefont
  {DiVincenzo}},\ }\href {\doibase 10.1103/PhysRevA.51.1015} {\bibfield
  {journal} {\bibinfo  {journal} {Physical Review A}\ }\textbf {\bibinfo
  {volume} {51}},\ \bibinfo {pages} {1015} (\bibinfo {year}
  {1995})}\BibitemShut {NoStop}%
\bibitem [{\citenamefont {Lloyd}(1995)}]{lloyd_almost_1995}%
  \BibitemOpen
  \bibfield  {author} {\bibinfo {author} {\bibfnamefont {S.}~\bibnamefont
  {Lloyd}},\ }\href {\doibase 10.1103/PhysRevLett.75.346} {\bibfield  {journal}
  {\bibinfo  {journal} {Physical Review Letters}\ }\textbf {\bibinfo {volume}
  {75}},\ \bibinfo {pages} {346} (\bibinfo {year} {1995})}\BibitemShut
  {NoStop}%
\bibitem [{\citenamefont {Barenco}\ \emph {et~al.}(1995)\citenamefont
  {Barenco}, \citenamefont {Bennett}, \citenamefont {Cleve}, \citenamefont
  {DiVincenzo}, \citenamefont {Margolus}, \citenamefont {Shor}, \citenamefont
  {Sleator}, \citenamefont {Smolin},\ and\ \citenamefont
  {Weinfurter}}]{barenco_elementary_1995}%
  \BibitemOpen
  \bibfield  {author} {\bibinfo {author} {\bibfnamefont {A.}~\bibnamefont
  {Barenco}}, \bibinfo {author} {\bibfnamefont {C.~H.}\ \bibnamefont
  {Bennett}}, \bibinfo {author} {\bibfnamefont {R.}~\bibnamefont {Cleve}},
  \bibinfo {author} {\bibfnamefont {D.~P.}\ \bibnamefont {DiVincenzo}},
  \bibinfo {author} {\bibfnamefont {N.}~\bibnamefont {Margolus}}, \bibinfo
  {author} {\bibfnamefont {P.}~\bibnamefont {Shor}}, \bibinfo {author}
  {\bibfnamefont {T.}~\bibnamefont {Sleator}}, \bibinfo {author} {\bibfnamefont
  {J.~A.}\ \bibnamefont {Smolin}}, \ and\ \bibinfo {author} {\bibfnamefont
  {H.}~\bibnamefont {Weinfurter}},\ }\href {\doibase 10.1103/PhysRevA.52.3457}
  {\bibfield  {journal} {\bibinfo  {journal} {Physical Review A}\ }\textbf
  {\bibinfo {volume} {52}},\ \bibinfo {pages} {3457} (\bibinfo {year}
  {1995})}\BibitemShut {NoStop}%
\bibitem [{imp()}]{implemented}%
  \BibitemOpen
  \href@noop {} {}\bibinfo {note} {K. Yamaguchi, L. Rullk\"otter, C. Tutschku, S. Wagner, I.~Shehzad, and A. Kempf, in preparation.}\BibitemShut {Stop}%
\bibitem [{\citenamefont {Kent}(2012)}]{kent_quantum_2012}%
  \BibitemOpen
  \bibfield  {author} {\bibinfo {author} {\bibfnamefont {A.}~\bibnamefont
  {Kent}},\ }\href {\doibase 10.1088/0264-9381/29/22/224013} {\bibfield
  {journal} {\bibinfo  {journal} {Classical and Quantum Gravity}\ }\textbf
  {\bibinfo {volume} {29}},\ \bibinfo {pages} {224013} (\bibinfo {year}
  {2012})}\BibitemShut {NoStop}%
\bibitem [{\citenamefont {Kent}(2013)}]{kent_no-summoning_2013}%
  \BibitemOpen
  \bibfield  {author} {\bibinfo {author} {\bibfnamefont {A.}~\bibnamefont
  {Kent}},\ }\href {\doibase 10.1007/s11128-012-0431-6} {\bibfield  {journal}
  {\bibinfo  {journal} {Quantum Information Processing}\ }\textbf {\bibinfo
  {volume} {12}},\ \bibinfo {pages} {1023} (\bibinfo {year}
  {2013})}\BibitemShut {NoStop}%
\bibitem [{\citenamefont {Hayden}\ and\ \citenamefont
  {May}(2016)}]{hayden_summoning_2016}%
  \BibitemOpen
  \bibfield  {author} {\bibinfo {author} {\bibfnamefont {P.}~\bibnamefont
  {Hayden}}\ and\ \bibinfo {author} {\bibfnamefont {A.}~\bibnamefont {May}},\
  }\href {\doibase 10.1088/1751-8113/49/17/175304} {\bibfield  {journal}
  {\bibinfo  {journal} {Journal of Physics A: Mathematical and Theoretical}\
  }\textbf {\bibinfo {volume} {49}},\ \bibinfo {pages} {175304} (\bibinfo
  {year} {2016})}\BibitemShut {NoStop}%
\bibitem [{\citenamefont {Adlam}\ and\ \citenamefont
  {Kent}(2016)}]{adlam_quantum_2016}%
  \BibitemOpen
  \bibfield  {author} {\bibinfo {author} {\bibfnamefont {E.}~\bibnamefont
  {Adlam}}\ and\ \bibinfo {author} {\bibfnamefont {A.}~\bibnamefont {Kent}},\
  }\href {\doibase 10.1103/PhysRevA.93.062327} {\bibfield  {journal} {\bibinfo
  {journal} {Physical Review A}\ }\textbf {\bibinfo {volume} {93}},\ \bibinfo
  {pages} {062327} (\bibinfo {year} {2016})}\BibitemShut {NoStop}%
\bibitem [{\citenamefont {Cleve}\ \emph {et~al.}(1999)\citenamefont {Cleve},
  \citenamefont {Gottesman},\ and\ \citenamefont {Lo}}]{cleve_how_1999}%
  \BibitemOpen
  \bibfield  {author} {\bibinfo {author} {\bibfnamefont {R.}~\bibnamefont
  {Cleve}}, \bibinfo {author} {\bibfnamefont {D.}~\bibnamefont {Gottesman}}, \
  and\ \bibinfo {author} {\bibfnamefont {H.-K.}\ \bibnamefont {Lo}},\ }\href
  {\doibase 10.1103/PhysRevLett.83.648} {\bibfield  {journal} {\bibinfo
  {journal} {Physical Review Letters}\ }\textbf {\bibinfo {volume} {83}},\
  \bibinfo {pages} {648} (\bibinfo {year} {1999})}\BibitemShut {NoStop}%
\bibitem [{\citenamefont {Gottesman}(2000)}]{gottesman_theory_2000}%
  \BibitemOpen
  \bibfield  {author} {\bibinfo {author} {\bibfnamefont {D.}~\bibnamefont
  {Gottesman}},\ }\href {\doibase 10.1103/PhysRevA.61.042311} {\bibfield
  {journal} {\bibinfo  {journal} {Physical Review A}\ }\textbf {\bibinfo
  {volume} {61}},\ \bibinfo {pages} {042311} (\bibinfo {year}
  {2000})}\BibitemShut {NoStop}%
\bibitem [{\citenamefont {Coffman}\ \emph {et~al.}(2000)\citenamefont
  {Coffman}, \citenamefont {Kundu},\ and\ \citenamefont
  {Wootters}}]{coffman_distributed_2000}%
  \BibitemOpen
  \bibfield  {author} {\bibinfo {author} {\bibfnamefont {V.}~\bibnamefont
  {Coffman}}, \bibinfo {author} {\bibfnamefont {J.}~\bibnamefont {Kundu}}, \
  and\ \bibinfo {author} {\bibfnamefont {W.~K.}\ \bibnamefont {Wootters}},\
  }\href {\doibase 10.1103/PhysRevA.61.052306} {\bibfield  {journal} {\bibinfo
  {journal} {Physical Review A}\ }\textbf {\bibinfo {volume} {61}},\ \bibinfo
  {pages} {052306} (\bibinfo {year} {2000})}\BibitemShut {NoStop}%
\bibitem [{\citenamefont {Knill}\ and\ \citenamefont
  {Laflamme}(1997)}]{knill_theory_1997}%
  \BibitemOpen
  \bibfield  {author} {\bibinfo {author} {\bibfnamefont {E.}~\bibnamefont
  {Knill}}\ and\ \bibinfo {author} {\bibfnamefont {R.}~\bibnamefont
  {Laflamme}},\ }\href {\doibase 10.1103/PhysRevA.55.900} {\bibfield  {journal}
  {\bibinfo  {journal} {Physical Review A}\ }\textbf {\bibinfo {volume} {55}},\
  \bibinfo {pages} {900} (\bibinfo {year} {1997})}\BibitemShut {NoStop}%
\bibitem [{\citenamefont {Roffe}(2019)}]{roffe_quantum_2019}%
  \BibitemOpen
  \bibfield  {author} {\bibinfo {author} {\bibfnamefont {J.}~\bibnamefont
  {Roffe}},\ }\href {\doibase 10.1080/00107514.2019.1667078} {\bibfield
  {journal} {\bibinfo  {journal} {Contemporary Physics}\ }\textbf {\bibinfo
  {volume} {60}},\ \bibinfo {pages} {226} (\bibinfo {year} {2019})}\BibitemShut
  {NoStop}%
\bibitem [{\citenamefont {Brun}(2020)}]{brun_quantum_2020}%
  \BibitemOpen
  \bibfield  {author} {\bibinfo {author} {\bibfnamefont {T.~A.}\ \bibnamefont
  {Brun}},\ }in\ \href {\doibase 10.1093/acrefore/9780190871994.013.35} {\emph
  {\bibinfo {booktitle} {Oxford {Research} {Encyclopedia} of {Physics}}}}\
  (\bibinfo {year} {2020})\BibitemShut {NoStop}%
\bibitem [{\citenamefont {Georgescu}(2020)}]{georgescu_25_2020}%
  \BibitemOpen
  \bibfield  {author} {\bibinfo {author} {\bibfnamefont {I.}~\bibnamefont
  {Georgescu}},\ }\href {\doibase 10.1038/s42254-020-0244-y} {\bibfield
  {journal} {\bibinfo  {journal} {Nature Reviews Physics}\ }\textbf {\bibinfo
  {volume} {2}},\ \bibinfo {pages} {519} (\bibinfo {year} {2020})}\BibitemShut
  {NoStop}%
\bibitem [{\citenamefont {Haus}\ and\ \citenamefont
  {Mullen}(1962)}]{haus_quantum_1962}%
  \BibitemOpen
  \bibfield  {author} {\bibinfo {author} {\bibfnamefont {H.~A.}\ \bibnamefont
  {Haus}}\ and\ \bibinfo {author} {\bibfnamefont {J.~A.}\ \bibnamefont
  {Mullen}},\ }\href {\doibase 10.1103/PhysRev.128.2407} {\bibfield  {journal}
  {\bibinfo  {journal} {Physical Review}\ }\textbf {\bibinfo {volume} {128}},\
  \bibinfo {pages} {2407} (\bibinfo {year} {1962})}\BibitemShut {NoStop}%
\bibitem [{\citenamefont {Caves}(1982)}]{caves_quantum_1982}%
  \BibitemOpen
  \bibfield  {author} {\bibinfo {author} {\bibfnamefont {C.~M.}\ \bibnamefont
  {Caves}},\ }\href {\doibase 10.1103/PhysRevD.26.1817} {\bibfield  {journal}
  {\bibinfo  {journal} {Physical Review D}\ }\textbf {\bibinfo {volume} {26}},\
  \bibinfo {pages} {1817} (\bibinfo {year} {1982})}\BibitemShut {NoStop}%
\bibitem [{\citenamefont {Nielsen}\ and\ \citenamefont
  {Chuang}(1997)}]{nielsen1997programmable}%
  \BibitemOpen
  \bibfield  {author} {\bibinfo {author} {\bibfnamefont {M.~A.}\ \bibnamefont
  {Nielsen}}\ and\ \bibinfo {author} {\bibfnamefont {I.~L.}\ \bibnamefont
  {Chuang}},\ }\href {\doibase 10.1103/PhysRevLett.79.321} {\bibfield
  {journal} {\bibinfo  {journal} {Physical Review Letters}\ }\textbf {\bibinfo
  {volume} {79}},\ \bibinfo {pages} {321} (\bibinfo {year} {1997})}\BibitemShut
  {NoStop}%
\bibitem [{\citenamefont {Yang}\ \emph {et~al.}(2020)\citenamefont {Yang},
  \citenamefont {Renner},\ and\ \citenamefont {Chiribella}}]{yang2020optimal}%
  \BibitemOpen
  \bibfield  {author} {\bibinfo {author} {\bibfnamefont {Y.}~\bibnamefont
  {Yang}}, \bibinfo {author} {\bibfnamefont {R.}~\bibnamefont {Renner}}, \ and\
  \bibinfo {author} {\bibfnamefont {G.}~\bibnamefont {Chiribella}},\ }\href
  {\doibase 10.1103/PhysRevLett.125.210501} {\bibfield  {journal} {\bibinfo
  {journal} {Physical review letters}\ }\textbf {\bibinfo {volume} {125}},\
  \bibinfo {pages} {210501} (\bibinfo {year} {2020})}\BibitemShut {NoStop}%
\bibitem [{\citenamefont {Lloyd}(2008)}]{lloyd_enhanced_2008}%
  \BibitemOpen
  \bibfield  {author} {\bibinfo {author} {\bibfnamefont {S.}~\bibnamefont
  {Lloyd}},\ }\href {\doibase 10.1126/science.1160627} {\bibfield  {journal}
  {\bibinfo  {journal} {Science}\ }\textbf {\bibinfo {volume} {321}},\ \bibinfo
  {pages} {1463} (\bibinfo {year} {2008})}\BibitemShut {NoStop}%
\bibitem [{\citenamefont {Barnum}\ \emph {et~al.}(2000)\citenamefont {Barnum},
  \citenamefont {Knill},\ and\ \citenamefont {Nielsen}}]{barnum_quantum_2000}%
  \BibitemOpen
  \bibfield  {author} {\bibinfo {author} {\bibfnamefont {H.}~\bibnamefont
  {Barnum}}, \bibinfo {author} {\bibfnamefont {E.}~\bibnamefont {Knill}}, \
  and\ \bibinfo {author} {\bibfnamefont {M.}~\bibnamefont {Nielsen}},\ }\href
  {\doibase 10.1109/18.850671} {\bibfield  {journal} {\bibinfo  {journal} {IEEE
  Transactions on Information Theory}\ }\textbf {\bibinfo {volume} {46}},\
  \bibinfo {pages} {1317} (\bibinfo {year} {2000})}\BibitemShut {NoStop}%
\bibitem [{\citenamefont {Devetak}(2005)}]{devetak_private_2005}%
  \BibitemOpen
  \bibfield  {author} {\bibinfo {author} {\bibfnamefont {I.}~\bibnamefont
  {Devetak}},\ }\href {\doibase 10.1109/TIT.2004.839515} {\bibfield  {journal}
  {\bibinfo  {journal} {IEEE Transactions on Information Theory}\ }\textbf
  {\bibinfo {volume} {51}},\ \bibinfo {pages} {44} (\bibinfo {year}
  {2005})}\BibitemShut {NoStop}%
\bibitem [{\citenamefont {Lloyd}(1997)}]{lloyd_capacity_1997}%
  \BibitemOpen
  \bibfield  {author} {\bibinfo {author} {\bibfnamefont {S.}~\bibnamefont
  {Lloyd}},\ }\href {\doibase 10.1103/PhysRevA.55.1613} {\bibfield  {journal}
  {\bibinfo  {journal} {Physical Review A}\ }\textbf {\bibinfo {volume} {55}},\
  \bibinfo {pages} {1613} (\bibinfo {year} {1997})}\BibitemShut {NoStop}%
\bibitem [{\citenamefont {Shor}(2002)}]{shor_quantum_2002}%
  \BibitemOpen
  \bibfield  {author} {\bibinfo {author} {\bibfnamefont {P.}~\bibnamefont
  {Shor}},\ }\href {https://www.msri.org/workshops/203/schedules/1181}
  {\bibfield  {journal} {\bibinfo  {journal} {In Lecture Notes, MSRI Workshop
  on Quantum Computation}\ } (\bibinfo {year} {2002})}\BibitemShut {NoStop}%
\bibitem [{\citenamefont {Schumacher}\ and\ \citenamefont
  {Nielsen}(1996)}]{schumacher_quantum_1996}%
  \BibitemOpen
  \bibfield  {author} {\bibinfo {author} {\bibfnamefont {B.}~\bibnamefont
  {Schumacher}}\ and\ \bibinfo {author} {\bibfnamefont {M.~A.}\ \bibnamefont
  {Nielsen}},\ }\href {\doibase 10.1103/PhysRevA.54.2629} {\bibfield  {journal}
  {\bibinfo  {journal} {Physical Review A}\ }\textbf {\bibinfo {volume} {54}},\
  \bibinfo {pages} {2629} (\bibinfo {year} {1996})}\BibitemShut {NoStop}%
\bibitem [{\citenamefont {Wilde}(2013)}]{wilde_quantum_2013}%
  \BibitemOpen
  \bibfield  {author} {\bibinfo {author} {\bibfnamefont {M.~M.}\ \bibnamefont
  {Wilde}},\ }\href {\doibase 10.1017/CBO9781139525343} {\emph {\bibinfo
  {title} {Quantum {Information} {Theory}}}}\ (\bibinfo  {publisher} {Cambridge
  University Press},\ \bibinfo {address} {Cambridge},\ \bibinfo {year}
  {2013})\BibitemShut {NoStop}%
\bibitem [{\citenamefont {Sleator}\ and\ \citenamefont
  {Weinfurter}(1995)}]{sleator_realizable_1995}%
  \BibitemOpen
  \bibfield  {author} {\bibinfo {author} {\bibfnamefont {T.}~\bibnamefont
  {Sleator}}\ and\ \bibinfo {author} {\bibfnamefont {H.}~\bibnamefont
  {Weinfurter}},\ }\href {\doibase 10.1103/PhysRevLett.74.4087} {\bibfield
  {journal} {\bibinfo  {journal} {Physical Review Letters}\ }\textbf {\bibinfo
  {volume} {74}},\ \bibinfo {pages} {4087} (\bibinfo {year}
  {1995})}\BibitemShut {NoStop}%
\bibitem [{\citenamefont {Nielsen}\ and\ \citenamefont
  {Chuang}(2010)}]{nielsen_quantum_2010}%
  \BibitemOpen
  \bibfield  {author} {\bibinfo {author} {\bibfnamefont {M.~A.}\ \bibnamefont
  {Nielsen}}\ and\ \bibinfo {author} {\bibfnamefont {I.~L.}\ \bibnamefont
  {Chuang}},\ }\href {https://doi.org/10.1017/CBO9780511976667} {\enquote
  {\bibinfo {title} {Quantum {Computation} and {Quantum} {Information}: 10th
  {Anniversary} {Edition}},}\ } (\bibinfo {year} {2010})\BibitemShut {NoStop}%
\bibitem [{\citenamefont {Yu}\ \emph {et~al.}(2013)\citenamefont {Yu},
  \citenamefont {Duan},\ and\ \citenamefont {Ying}}]{yu_five_2013}%
  \BibitemOpen
  \bibfield  {author} {\bibinfo {author} {\bibfnamefont {N.}~\bibnamefont
  {Yu}}, \bibinfo {author} {\bibfnamefont {R.}~\bibnamefont {Duan}}, \ and\
  \bibinfo {author} {\bibfnamefont {M.}~\bibnamefont {Ying}},\ }\href {\doibase
  10.1103/PhysRevA.88.010304} {\bibfield  {journal} {\bibinfo  {journal}
  {Physical Review A}\ }\textbf {\bibinfo {volume} {88}},\ \bibinfo {pages}
  {010304} (\bibinfo {year} {2013})}\BibitemShut {NoStop}%
\end{thebibliography}
%

\let\addcontentsline\oldaddcontentsline

\appendix

\onecolumngrid
\begin{center}
{\large \bf End Matter}\\
\end{center}
\twocolumngrid

\begin{center}
\textbf{Appendix A: Variants of the protocol}    
\end{center}
Let us consider the scenario where the owner of the quantum data, Alice, loses access to some of the $n$ noise qubits but retains access to at least one noise qubit. In this case, the owner can still recover the original state of $A$ from any subsystem composed of one pair of signal and noise qubits plus at least one half of each of the remaining $(n-1)$ pairs of signal and noise qubits. For example, the original state of $A$ can be recovered using $\s_1\s_2$ and $\n_1\n_3\n_4\cdots \n_n$ by replacing $\sigma_\mu^{(\n_2)\top}$ with $\sigma_\mu^{(\s_2)}$ in Eq.~\eqref{eq:decoding_op}.  

While, therefore, the loss of some noise or signal qubits can be compensated, the loss of even just one pair of signal and noise qubits makes it impossible to retrieve qubit $A$, and the channel capacity drops to zero. This is because the reduced state of $(n-1)$ pairs of signal and noise qubits is given by $\frac{1}{4}\sum_{\mu=0}^3\ket{\phi_\mu}\bra{\phi_\mu}^{\otimes (n-1)}$ after the encoding, independently of $\ket{\psi}_A$, as seen in Eq.~\eqref{eq:state_after_encoding}. 

However, in the case where Alice, the owner of the quantum data, is also the owner of the quantum clouds, or is trusting the quantum cloud provider(s), the owner can enable an additional security feature that allows her to recover the original qubit $A$ even in the case of a loss of all noise qubits. 
To enable this security feature, the owner sends not only the $n$ copies to quantum clouds but also sends the post-encoding qubit $A$ (which is maximally mixed) to a quantum cloud. Now even if Alice were to lose all of her noise qubits, she could still recover the original qubit $A$, namely by running the encoding unitary in reverse on the collection of all signal qubits and the post-encoding qubit $A$.  
In addition, if Alice maintains access to all $n$ noise qubits, she can recover the original qubit $A$ by using any one of the signal qubits, or also by using the post-encoding qubit $A$, at least if $n$ is even, as we show in the Supplementary Material~\cite{sm}. 
In effect, at least if $n$ is even, the post-encoding qubit $A$ can serve as an encrypted clone as well.

Finally, among possible variants, let us mention the
possibility to create a large number of encrypted clones
of $A$ by repeatedly creating say two additional clones of $A$, by using 
the $n = 2$ encrypted cloning method. 
In this case, while the last created pair of encrypted clones requires all noise qubits for decryption, the first pair of encrypted clones requires only the first two noise qubits for decryption. 

Alternatively, a favorable scaling of the number of noise qubits required for decryption can be obtained, for example, by creating $3$ encrypted clones (counting also $A$ as an encrypted clone) using the $n=2$ method, then creating $3$ encrypted clones of each of these encrypted clones and so on, say $k$ times. This yields $3^k$ encrypted clones along with $2(1+3+\cdots+3^{k-1})=3^k-1$ noise qubits. 
This means that to create $m+1:=3^k$ encrypted clones, $m$ noise qubits are required, as in the original method for $n=m$ (counting also $A$ as an encrypted clone).
However, when using this iterated encrypted cloning method, each of the $3^k$ encrypted clone requires for its decryption merely $2k$ specific noise qubits.

\widetext

\clearpage

\setcounter{page}{1}
\begin{center}
{\large \bf Supplemental Material for\\
``\inserttitle''}\\
\vspace*{0.3cm}
Koji Yamaguchi$^{1}$ and Achim Kempf$^{2,3,4,5}$\\
\vspace*{0.1cm}
$^{1}${\small \it Department of Communication Engineering and Informatics,
University of Electro-Communications,\\
1-5-1 Chofugaoka, Chofu, Tokyo, 182-8585, Japan}

$^{2}${\small \it Department of Applied Mathematics, University of Waterloo, Waterloo, Ontario, N2L
3G1, Canada}
\\
$^{3}${\small \it Department of Physics, University of Waterloo, Waterloo, ON N2L 3G1, Canada}
\\
$^{4}${\small \it Institute for Quantum Computing, University of Waterloo, Waterloo, ON N2L 3G1, Canada}
\\
$^{5}${\small \it Perimeter Institute for Theoretical Physics, Waterloo, Ontario N2L 2Y5, Canada}
\end{center}

\renewcommand{\theequation}{S.\arabic{equation}}
\setcounter{equation}{0}

\tableofcontents

\section{Alternative proof of the reconstructability of qubit $A$}

\noindent We begin by recalling the precise definition of the quantum capacity of a quantum channel \cite{barnum_quantum_2000,devetak_private_2005}. 
To this end, let $\mathcal{B}(\mathcal{H})$ denote the set of all bounded linear operators on a Hilbert space $\mathcal{H}$. For a quantum channel $\mathcal{N}:\mathcal{B}(\mathcal{H}_A)\to\mathcal{B}(\mathcal{H}_B)$, a coding scheme is a sequence of quantum channels $\{(\mathcal{E}^{(n)},\mathcal{D}^{(n)})\}_{n=1}^\infty$ such that
\begin{align}
    &\mathcal{E}^{(n)}:\mathcal{B}(\mathcal{H}_s^{\otimes n})\to \mathcal{B}(\mathcal{H}_A^{\otimes n}),\quad \mathcal{D}^{(n)}:\mathcal{B}(\mathcal{H}_B^{\otimes n})\to \mathcal{B}(\mathcal{H}_s^{\otimes n}),
\end{align}
where $\mathcal{H}_s$ denotes a Hilbert space for a quantum source. 
For a quantum channel $\mathcal{A}:\mathcal{B}(\mathcal{H}')\to\mathcal{B}(\mathcal{H}')$ and a sub-Hilbert space $\mathcal{H}$ of $\mathcal{H}'$, we define the pure-state fidelity by
\begin{align}
    \mathcal{F}_p\left(\mathcal{H},\mathcal{A}\right)\coloneqq \min_{\ket{\psi}\in\mathcal{H}}\braket{\psi|\mathcal{A}(\ket{\psi}\bra{\psi})|\psi},
\end{align}
where the minimization is taken over the set of all pure states in $\mathcal{H}$. We say that a rate $R>0$ is achievable with a channel $\mathcal{N}$ if there exist a sequence of subspace $\mathcal{H}^{(n)}$ of $\mathcal{H}_{s}^{\otimes n}$ and a coding scheme $(\mathcal{E}^{(n)},\mathcal{D}^{(n)})$ such that 
\begin{align}
    \lim_{n\to\infty}\mathcal{F}_p\left(\mathcal{H}^{(n)},\mathcal{D}^{(n)}\circ \mathcal{N}^{\otimes n}\circ\mathcal{E}^{(n)}\right)=1
\end{align}
and 
\begin{align}
    \limsup_{n\to\infty}\frac{1}{n}\log_2\dim\left(\mathcal{H}^{(n)}\right)=R.
\end{align}
The quantum capacity $C_Q(\mathcal{N})$ of a quantum channel $\mathcal{N}$ is defined as
\begin{align}
    C_Q(\mathcal{N})\coloneqq \sup\left\{R>0\,\middle|\, R \text{ is achievable with }\mathcal{N}\right\}. 
\end{align}

\noindent We will now briefly review the relation between quantum capacity and coherent information, based on the Lloyd-Shor-Devetak (LSD) theorem \cite{lloyd_capacity_1997,shor_quantum_2002, devetak_private_2005}. To this end, let us recall that the coherent information for a bipartite state $\rho_{\widetilde{A}A}$ is defined \cite{schumacher_quantum_1996} by
\begin{align}
    I(\widetilde{A}\rangle A)_\rho\coloneqq S_{\mathrm{vN}}(\rho_A)-S_{\mathrm{vN}}(\rho_{\widetilde{A}A}),
\end{align}
where $\rho_A\coloneqq \mathrm{Tr}_{\tilde{A}}(\rho_{\tilde{A}A})$ and $S_{\mathrm{vN}}(\rho)\coloneqq -\mathrm{Tr}(\rho\log_2(\rho))$ is the von Neumann entropy. The coherent information of a channel $\mathcal{N}$ is defined as
\begin{align}
    Q\left(\mathcal{N}\right)\coloneqq \max_{\phi_{\widetilde{A}A}}I(\widetilde{A}\rangle B)_\rho,\label{eq:coh_info_channel}
\end{align}
where the maximization is taken over the set of all bipartite pure states $\phi_{\widetilde{A}A}$ and $\rho\coloneqq \mathcal{I}_{\widetilde{A}}\otimes \mathcal{N}_{A\to B}(\phi_{\widetilde{A}A})$. Here, the subscript of the channel $\mathcal{N}_{A\to B}$ indicates its input system $A$ and output system $B$. The LSD theorem \cite{lloyd_capacity_1997,shor_quantum_2002, devetak_private_2005} asserts
\begin{align}
    C_Q(\mathcal{N})= \lim_{k\to\infty}\frac{1}{k}Q\left(\mathcal{N}^{\otimes k}\right).\label{eq:sld_formula}
\end{align}
See, e.g., \cite{wilde_quantum_2013} for further detail.

It is typically difficult to calculate the quantum capacity because the right-hand side of Eq.~\eqref{eq:sld_formula} requires optimization for an infinitely large $k$. There are several examples for which the exact value of the quantum capacity is known. For example, the quantum capacity vanishes for any anti-degradable channel \cite{caruso_degradability_2006,bennett_capacities_1997,giovannetti_broadband_2003,giovannetti_entanglement_2003,holevo_entanglement-breaking_2008} because of the no-cloning constraint as mentioned in the main text. For a generic channel, it is common to analyze the bounds of the quantum capacity. From the LSD theorem and Eq.~\eqref{eq:coh_info_channel}, $C_Q(\mathcal{N})\geq Q(\mathcal{N})$ holds for any quantum channel $\mathcal{N}$. From the definition in Eq.~\eqref{eq:coh_info_channel}, we further find
\begin{align}
    C_Q(\mathcal{N})\geq I(\widetilde{A}\rangle B)_\rho\label{eq:lowerbound_quantum_capacity}
\end{align}
holds for $\rho\coloneqq \mathcal{I}_{\widetilde{A}}\otimes \mathcal{N}_{A\to B}(\phi_{\widetilde{A}A})$, where $\phi_{\widetilde{A}A}$ is an arbitrary pure state. In the next section, we will use this inequality to derive the lower bound of the quantum capacity in our model. 
\medskip\newline

\noindent We will now provide an alternative proof for the reconstructability of qubit $A$, namely by showing that the coherent information, which provides a lower bound for the quantum capacity, is 1.

Concretely, we derive a lower bound of the quantum capacity for encrypted cloning with the encoding operation 
\begin{align}
    U_{\mathrm{enc}}^{(n)}(t)\coloneqq e^{-\ii t \sigma_{1}^{(A)}\otimes \left(\bigotimes_{i=1}^n\sigma_{1}^{(\s_i)}\right)}e^{-\ii t \sigma_{3}^{(A)}\otimes \left(\bigotimes_{i=1}^n\sigma_{3}^{(\s_i)}\right)}\label{eq:encoding_op_t}
\end{align}
for $t\in\mathbb{R}$ by explicitly calculating the coherent information. Note that the encoding operation in Eq.~\eqref{eq:encoding_optimal} in the main text is a special case where $t=\pi/4$. We show the coherent information for $\rho\coloneqq \mathcal{I}_{\widetilde{A}}\otimes \mathcal{N}_{A\to \s_1\n_1\n_2\cdots \n_n}(\phi_{\widetilde{A}A})$,  $\phi_{\widetilde{A}A}\coloneqq \ket{\phi}\bra{\phi}_{\widetilde{A}A}$ and $\ket{\phi}_{\widetilde{A}A}\coloneqq (\ket{0}_{\widetilde{A}}\ket{0}_{A}+\ket{1}_{\widetilde{A}}\ket{1}_{A})/\sqrt{2}$ is given by
\begin{align}
    I(\widetilde{A}\rangle  \s_1\n_1\n_2\cdots \n_n)_\rho 
    &=-\sum_{\mu=0}^3\lambda_\mu(t)\log_2\lambda_\mu(t)-1,\label{eq:coherent_information_result}
\end{align}
where $\lambda_0(t)\coloneqq \cos^4t$, $\lambda_1(t)=\lambda_3(t)\coloneqq \sin^2{t}\cos^2{t}$ and $\lambda_2(t)\coloneqq \sin^4t$. Since $I(\widetilde{A}\rangle  \s_1\n_1\n_2\cdots \n_n)_\rho =1$ for $t=\pi/4$, Eq.~\eqref{eq:lowerbound_quantum_capacity} implies $C_Q(\mathcal{N}^{(n)}_{A\to \s_1\n_1\n_2\cdots \n_n})|_{t=\frac{\pi}{4}}=1$, providing an alternative proof of Eq.\eqref{eq:channel_capacity_copy}.

In order to calculate the coherent information, let us first expand the encoding operation as
\begin{align}
    U_{\mathrm{enc}}^{(n)}(t)=\sum_{\mu=0}^3c_\mu(t) \Sigma_\mu,\quad \Sigma_\mu&\coloneqq \sigma_\mu^{(A)}\otimes\left(\bigotimes_{i=1}^n \sigma_\mu^{(\s_i)}\otimes \mathbb{I}^{(\n_i)}\right),
\end{align}
where we used $\sigma_{1}^{\otimes n}\sigma_{3}^{\otimes n}=(-\ii)^n \sigma_2^{\otimes n}$ and defined $c_0(t)\coloneqq \cos^2{t}$, $c_1(t)=c_3(t)\coloneqq-\ii\cos{t}\sin{t}$ and $c_2(t)\coloneqq -(-\ii)^{n+1}\sin^2{t}$. After the encoding, the total system evolves into
\begin{align}
    U_{\mathrm{enc}}^{(n)}(t)\ket{\phi}_{\widetilde{A}A}\left(\bigotimes_{i=1}^n\ket{\phi}_{\s_i\n_i}\right)=\sum_{\mu=0}^3 c_\mu (t)\ket{\Phi_\mu}_{\widetilde{A}A\s_1\n_1\cdots \s_n\n_n},
\end{align}
where 
\begin{align}
    \ket{\Phi_\mu}_{\widetilde{A}\s_1\n_1\cdots \s_n\n_n}\coloneqq \mathbb{I}_{\widetilde{A}}\otimes \Sigma_\mu\ket{\phi}_{\widetilde{A}A}\left(\bigotimes_{i=1}^n\ket{\phi}_{\s_i\n_i}\right).
\end{align}

Now, we calculate the eigenvalues of the reduced state
\begin{align}
    &\rho_{\widetilde{A}\s_1\n_1\n_2\cdots \n_n}(t)=\sum_{\mu,\nu=0}^3 c_\mu (t)c_\nu^*(t)\mathrm{Tr}_{A \s_2\s_3\cdots \s_n}\left(\ket{\Phi_\mu}\bra{\Phi_\nu}\right).
\end{align}
In the following, we use the fact that $O\otimes \mathbb{I}\ket{\phi}=\mathbb{I}\otimes O^{\top}\ket{\phi}$ holds for any linear operator $O$, where $\top$ implies a transpose operation with respect to the computational basis $\{\ket{0},\ket{1}\}$. 
With this trick, we can rewrite
\begin{align}
    \ket{\Phi_\mu}_{\widetilde{A}A\s_1\n_1\cdots \s_n\n_n}&=\sigma_{\mu}^{(\widetilde{A})\top}\otimes\mathbb{I}_A\otimes \left(\bigotimes_{i=1}^n\mathbb{I}^{(\s_i)}\otimes \sigma_\mu^{(\n_i)\top}\right) \ket{\phi}_{\widetilde{A}A}\ket{\phi}_{\s_1\n_1}\ket{\phi}_{\s_2\n_2}\cdots \ket{\phi}_{\s_n\n_n}.
\end{align}
Therefore, 
\begin{align}
    &\mathrm{Tr}_{A\s_2\s_3\cdots \s_n}\left(\ket{\Phi_\mu}\bra{\Phi_\nu}\right)=\frac{1}{2}\sigma_{\mu}^{(\widetilde{A})\top}\sigma_\nu^{(\widetilde{A})\top}\otimes \left(\sigma_{\mu}^{(\s_1)}\otimes \mathbb{I}^{(\n_1)}\phi_{\s_1\n_1}\sigma_{\nu}^{(\s_1)}\otimes \mathbb{I}^{(\n_1)}\right)\otimes \left(\bigotimes_{j=2}^n\frac{1}{2}\sigma_{\mu}^{(\n_j)\top}\sigma_\nu^{(\n_j)\top}\right).\label{eq:trace_phimu_phinu}
\end{align}
Let us introduce
\begin{align}
    V\coloneqq \sum_{\mu=0}^4\sigma_{\mu}^{(\widetilde{A})\top}\otimes \left(\sigma_{\mu}^{(\s_1)}\otimes \mathbb{I}^{(\n_1)}\phi_{\s_1\n_1}\sigma_{\mu}^{(\s_1)}\otimes \mathbb{I}^{(\n_1)}\right)\otimes \left(\bigotimes_{j=2}^n\sigma_{\mu}^{(\n_j)\top}\right).
\end{align}
This linear operator $V$ is unitary since $\{\sigma_{\mu}^{(\s_1)}\otimes \mathbb{I}^{(\n_1)}\ket{\phi}_{\s_1\n_1}\}_{\mu=0}^3$ is an orthonormal basis for a two-qubit system. By using this unitary operation, we have
\begin{align}
    &V\rho_{\widetilde{A}\s_1\n_1\n_2\cdots \n_n}(t) V^\dag=\frac{1}{2}\mathbb{I}^{(\widetilde{A})}\otimes \left(\sum_{\mu,\nu=0}^3c_\mu (t)c_\nu^*(t)\sigma_{\mu}^{(\s_1)}\otimes \mathbb{I}^{(\n_1)}\phi_{\s_1\n_1}\sigma_{\nu}^{(\s_1)}\otimes \mathbb{I}^{(\n_1)}\right)\otimes \left(\bigotimes_{j=2}^n\frac{1}{2}\mathbb{I}^{(\n_j)}\right).
\end{align}
Since $\sum_{\mu=0}^3|c_\mu(t)|^2=1$, we find that
\begin{align}
    &\left(\sum_{\mu,\nu=0}^3c_\mu (t)c_\nu^*(t)\sigma_{\mu}^{(\s_1)}\otimes \mathbb{I}^{(\n_1)}\phi_{\s_1\n_1}\sigma_{\nu}^{(\s_1)}\otimes \mathbb{I}^{(\n_1)}\right)^2
    =\sum_{\mu,\nu=0}^3c_\mu (t)c_\nu^*(t)\sigma_{\mu}^{(\s_1)}\otimes \mathbb{I}^{(\n_1)}\phi_{\s_1\n_1}\sigma_{\nu}^{(\s_1)}\otimes \mathbb{I}^{(\n_1)}
\end{align}
holds, implying that $\sum_{\mu,\nu=0}^3c_\mu (t)c_\nu^*(t)\sigma_{\mu}^{(\s_1)}\otimes \mathbb{I}^{(\n_1)}\phi_{\s_1\n_1}\sigma_{\nu}^{(\s_1)}\otimes \mathbb{I}^{(\n_1)}$ is a pure state. Therefore, the non-vanishing eigenvalues of $\rho_{\widetilde{A}\s_1\n_1\n_2\cdots \n_n}(t)$ are $1/2^n$ and hence the von Neumann entropy is calculated as
\begin{align}
    S_{\mathrm{vN}}(\rho_{\widetilde{A}\s_1\n_1\n_2\cdots \n_n}(t))=n
\end{align}
for any $t\in\mathbb{R}$. 

Next, we calculate the eigenvalues of $\rho_{\s_1 \n_1\n_2\cdots \n_n}(t)$. From Eq.~\eqref{eq:trace_phimu_phinu}, we get
\begin{align}
    &\mathrm{Tr}_{\widetilde{A}A\s_2\s_3\cdots \s_n}\left(\ket{\Phi_\mu}\bra{\Phi_\nu}\right)=\delta_{\mu\nu}\left(\sigma_{\mu}^{(\s_1)}\otimes \mathbb{I}^{(\n_1)}\phi_{\s_1\n_1}\sigma_{\nu}^{(\s_1)}\otimes \mathbb{I}^{(\n_1)}\right)\otimes \left(\bigotimes_{j=2}^n\frac{1}{2}\sigma_{\mu}^{(\n_j)\top}\sigma_\nu^{(\n_j)\top}\right).
\end{align}
Therefore, it holds
\begin{align}
    &\rho_{\s_1\n_1\n_2\cdots \n_n}(t)=\rho_{\s_1\n_1}(t)\otimes \left(\bigotimes_{j=2}^n\frac{1}{2}\mathbb{I}^{(\n_j)}\right),\quad \rho_{\s_1\n_1}(t)\coloneqq \sum_{\mu=0}^3|c_\mu(t)|^2\left(\sigma_{\mu}^{(\s_1)}\otimes \mathbb{I}^{(\n_1)}\phi_{\s_1\n_1}\sigma_{\mu}^{(\s_1)}\otimes \mathbb{I}^{(\n_1)}\right).
\end{align}
Since $\{\sigma_{\mu}^{(\s_1)}\otimes \mathbb{I}^{(\n_1)}\ket{\phi}_{\s_1\n_1}\}_{\mu=0}^3$ are orthonormal, the eigenvalues of $ \rho_{\s_1\n_1}(t)$ are given by $\{\lambda_\mu(t)\}_{\mu=0}^3$ with $\lambda_\mu(t)=|c_\mu(t)|^2$. Thus, we get
\begin{align}
    S_{\mathrm{vN}}(\rho_{\s_1\n_1\n_2\cdots \n_n}(t))&=-2^{n-1}\sum_{\mu=0}^3\frac{\lambda_\mu(t)}{2^{n-1}}\log_2{\frac{\lambda_\mu(t)}{2^{n-1}}}\nonumber\\
    &=n-1-\sum_{\mu=0}^3\lambda_\mu(t)\log_2{\lambda_\mu(t)}.
\end{align}

From the definition of coherent information, we finally get
\begin{align}
    I(\widetilde{A}\rangle \s_1\n_1\n_2\cdots \n_n)_\rho  &=S_{\mathrm{vN}}(\rho_{\s_1\n_1\n_2\cdots \n_n}(t))-S_{\mathrm{vN}}(\rho_{\widetilde{A}\s_1\n_1\n_2\cdots \n_n}(t))\nonumber\\
    &=-\sum_{\mu=0}^3\lambda_\mu(t)\log_2\lambda_\mu(t)-1,\label{eq:formula_coh_info}
\end{align}
which completes the proof of Eq.~\eqref{eq:coherent_information_result}. 

\begin{figure}[htbp]
    \centering
    \includegraphics{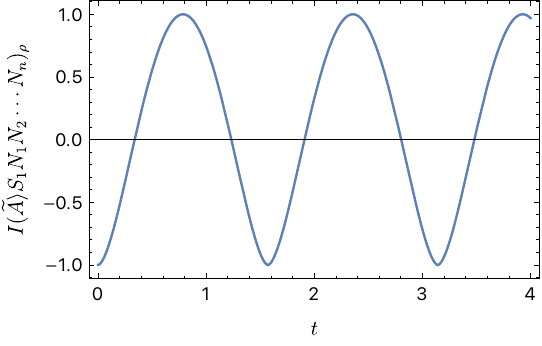}
    \caption{Plot for coherent information given in Eq.~\eqref{eq:formula_coh_info}. $I(\widetilde{A}\rangle \s_1\n_1\n_2\cdots \n_n)_\rho =1$ if $t=\frac{\pi}{4}+\frac{\pi}{2}m$ for $m\in\mathbb{Z}$.}
    \label{fig:plot_coh_info}
\end{figure}

Figure~\ref{fig:plot_coh_info} shows the plot for coherent information. In particular, when $t=\frac{\pi}{4}+\frac{\pi}{2}m$ for $m\in\mathbb{Z}$, $\lambda_\mu=\frac{1}{4}$ for $\mu=0,1,2,3$, implying that 
\begin{align}
    I(\widetilde{A}\rangle \s_1\n_1\n_2\cdots \n_n)_\rho \biggl|_{t=\frac{\pi}{4}+\frac{\pi}{2}m} =1.
\end{align}
Since $I(\widetilde{A}\rangle \s_1\n_1\n_2\cdots \n_n)_\rho $ is a lower bound of the quantum capacity $C_Q$, we get
\begin{align}
    C_Q\left(\mathcal{N}^{(n)}_{A\to \s_1\n_1\n_2\cdots \n_n}\right)\biggl|_{t=\frac{\pi}{4}+\frac{\pi}{2}m}=1,
\end{align}
which provides an alternative proof of Eq.~\eqref{eq:channel_capacity_copy}.

\section{Proof that the encryption is perfect}
Here we prove that, after encoding for $n>1$, no information about the original state of the qubit $A$ is stored locally in any signal qubit or the qubit $A$. 
Tracing over the qubit $A$ and a pair of the signal and noise qubit in Eq.~\eqref{eq:state_after_encoding}, the reduced state of $(n-1)$ pairs of signal and noise qubits is given by $\frac{1}{4}\sum_{\mu=0}^3\ket{\phi_\mu}\bra{\phi_\mu}^{\otimes (n-1)}$. Therefore, each signal qubit remains in the maximally mixed state after encoding. From Eq.~\eqref{eq:state_after_encoding}, the reduced state of qubit $A$ is given by
\begin{align}
    \rho_A=\frac{1}{4}\sum_{\mu=0}^3\sigma_\mu^{(A)}\ket{\psi}\bra{\psi}\sigma_\mu^{(A)}
\end{align}
for any integer $n$, where we have used $\braket{\phi_\mu|\phi_\nu}=\delta_{\mu\nu}$. From
\begin{align}
    \frac{1}{4}\sum_{\mu=0}^3\sigma_\mu\ket{0}\bra{0}\sigma_\mu&=\frac{1}{4}\left(\ket{0}\bra{0}+\ket{1}\bra{1}+\ket{1}\bra{1}+\ket{0}\bra{0}\right),\\
    \frac{1}{4}\sum_{\mu=0}^3\sigma_\mu\ket{1}\bra{1}\sigma_\mu&=\frac{1}{4}\left(\ket{1}\bra{1}+\ket{0}\bra{0}+\ket{0}\bra{0}+\ket{1}\bra{1}\right),\\
    \frac{1}{4}\sum_{\mu=0}^3\sigma_\mu\ket{0}\bra{1}\sigma_\mu
    &=\frac{1}{4}\left(\ket{0}\bra{1}+\ket{1}\bra{0}-\ket{1}\bra{0}-\ket{0}\bra{1}\right)=0,\\
    \frac{1}{4}\sum_{\mu=0}^3\sigma_\mu\ket{1}\bra{0}\sigma_\mu&=\left(\frac{1}{4}\sum_{\mu=0}^3\sigma_\mu\ket{0}\bra{1}\sigma_\mu\right)^\dag =0,
\end{align}
we find $\rho_A=\frac{\mathbb{I}}{2}$, regardless of the initial state of qubit $A$. 

Comment: It would be possible to extend the encrypted cloning method to the case $n=1$ by defining the decryption through: $U_{\mathrm{dec}}^{(1)}\coloneqq \sum_{\mu=0}^3\alpha_\mu\left(\ket{\phi_\mu}\bra{\phi_\mu}_{\s_1\n_1}\right)$. In this case, the encrypted clone $S_1$ is not fully encrypted, however, since as can readily be checked, its state retains a dependency on the state $\vert \psi\rangle_A$. 

\section{At least for even $n$, also qubit $A$ can serve as an encrypted clone.}
From Eq.~\eqref{eq:state_after_encoding}, the state of the total system after encoding can also be written as
\begin{align}
    U_{\mathrm{enc}}^{(n)}\left(\ket{\psi}_A\otimes\left(\bigotimes_{i=1}^n\ket{\phi}_{\s_i\n_i}\right)\right)
    &=\frac{1}{2}\sum_{\mu=0}^3\alpha_\mu^{-1} \sigma_\mu^{(A)}\otimes \left(\bigotimes_{i=1}^n\sigma_\mu^{(\s_i)}\right)\ket{\psi}_A\otimes\left(\bigotimes_{i=1}^n\ket{\phi}_{\s_i\n_i}\right)\\
    &=\frac{1}{2}\sum_{\mu=0}^3\alpha_\mu^{-1} \sigma_\mu^{(A)}\otimes \left(\bigotimes_{i=1}^n\sigma_\mu^{(\n_i)\top}\right)\ket{\psi}_A\otimes\left(\bigotimes_{i=1}^n\ket{\phi}_{\s_i\n_i}\right).
\end{align}
Since 
\begin{align}
   \sigma_{\mu}^{(\n_i)\top}=
   \begin{cases}
        \sigma_\mu^{(\n_i)}\quad &(\text{if }\mu=0,1,3)\\
       -\sigma_\mu^{(\n_i)}\quad &(\text{if }\mu=2)
   \end{cases},
\end{align}
we have $\left(\bigotimes_{i=1}^n\sigma_\mu^{(\n_i)\top}\right)=\left(\bigotimes_{i=1}^n\sigma_\mu^{(\n_i)}\right)$ for $\mu=0,1,2,3$ if $n$ is an even integer. Therefore, for even $n$, we get
\begin{align}
    U_{\mathrm{enc}}^{(n)}\left(\ket{\psi}_A\otimes\left(\bigotimes_{i=1}^n\ket{\phi}_{\s_i\n_i}\right)\right)&=U'_{AN_1\cdots N_n}\left(\ket{\psi}_A\otimes\left(\bigotimes_{i=1}^n\ket{\phi}_{\s_i\n_i}\right)\right),
\end{align}
where
\begin{align}
    U'_{AN_1\cdots N_n}&\coloneqq e^{-\frac{\pi\ii }{4} \sigma_{1}^{(A)}\otimes \left(\bigotimes_{i=1}^n\sigma_{1}^{(\n_i)}\right)}e^{-\frac{\pi\ii}{4} \sigma_{3}^{(A)}\otimes \left(\bigotimes_{i=1}^n\sigma_{3}^{(\n_i)}\right)}
\end{align}
is a unitary operator acting on qubits $AN_1\cdots N_n$, implying that we can decrypt the original state of qubit $A$ from qubit $A$ with the key $N_1\cdots N_n$ by performing $U^{\prime \dag}_{AN_1\cdots N_n}$.

\section{Decomposition of encryption and decryption operations into two-qubit gates}

The encryption and decryption operations in our model are described by unitary evolutions that involve multiple qubits. In practice, it is difficult to directly implement such a many-qubit interaction. We here provide an explicit decomposition of encryption and decryption operations into two-qubit gates, which would be helpful in their experimental realization. As a consequence, we find that the number of two-qubit gates needed to create $n$ encrypted clones, and to subsequently decrypt one of them, increases linearly with $n$.

First, let us investigate the factor $e^{-\ii t \sigma_3^{(A)}\otimes\left(\bigotimes_{i=1}^n\sigma_3^{(\s_i)}\right)}$ in the encryption operation in Eq.~\eqref{eq:encoding_op_t}. In the computational basis, this operation changes the phase by $e^{-\ii t}$ or $e^{\ii t}$ depending on the parity of the number of qubits involved. Therefore, it can be realized by the following quantum circuit, where we used a standard notation for rotation $R_i(\theta)\coloneqq e^{-\ii \frac{\theta}{2}\sigma_i}$:
\begin{center}
    \begin{quantikz}
        e^{-\ii t \sigma_3^{(A)}\otimes\left(\bigotimes_{i=1}^n\sigma_3^{(\s_i)}\right)}
    \end{quantikz}=
    \begin{quantikz}
        & \wire[l][1]["A"{above,pos=0.2}]{a} & \ctrl{1} & \qw & &\qw & \qw & \qw & \qw & \qw &\ctrl{1} & \qw & \qw\\
        &\wire[l][1]["\s_1"{above,pos=0.2}]{a} & \targ{} & \ctrl{1} & &\qw & \qw & \qw & \qw & \ctrl{1}& \targ{}&\qw& \qw\\
        & \wire[l][1]["\s_2"{above,pos=0.2}]{a}& & \targ{} &\wire[l][1]["\ddots"{below,pos=0.1}]{a} & & & &  \wire[l][1]["\iddots"{below,pos=0.1}]{a} &\targ{}& \qw &\qw& \qw \\
        &\wire[l][1]["\s_{n-1}"{above,pos=0.2}]{a} & \qw & \qw & \qw & \ctrl{1} & \qw & \ctrl{1} &  \qw &\qw &\qw& \qw& \qw \\
        & \wire[l][1]["\s_n"{above,pos=0.2}]{a} & \qw & \qw & \qw &\targ{} & \gate{R_z(2t)} & \targ{} &  \qw & \qw &\qw& \qw& \qw\\
    \end{quantikz}
\end{center}
Since the other part of the encryption operation, $e^{-\ii t \sigma_1^{(A)}\otimes\left(\bigotimes_{i=1}^n\sigma_1^{(\s_i)}\right)}$, is related to $e^{-\ii t \sigma_3^{(A)}\otimes \left(\bigotimes_{i=1}^n\sigma_3^{(\s_i)}\right)}$ via local unitary operations, it can be implemented as follows, where $H$ denotes the Hadamard gate:
\begin{center}
    \begin{quantikz}
        e^{-\ii t \sigma_1^{(A)}\otimes\left(\bigotimes_{i=1}^n\sigma_1^{(\s_i)}\right)}
    \end{quantikz}=
    \begin{quantikz}
        & \wire[l][1]["A"{above,pos=0.2}]{a} &\gate{H}&\ctrl{1} & \qw & &\qw & \qw & \qw & \qw & \qw &\ctrl{1} & \qw&\gate{H}&\qw\\
        &\wire[l][1]["\s_1"{above,pos=0.2}]{a} & \gate{H}&\targ{} & \ctrl{1} & &\qw & \qw & \qw & \qw & \ctrl{1}& \targ{}&\qw& \gate{H}&\qw\\
        & \wire[l][1]["\s_2"{above,pos=0.2}]{a}&\gate{H}& & \targ{} &\wire[l][1]["\ddots"{below,pos=0.1}]{a} & & & &  \wire[l][1]["\iddots"{below,pos=0.1}]{a} &\targ{}& \qw &\qw& \gate{H} &\qw\\
        &\wire[l][1]["\s_{n-1}"{above,pos=0.2}]{a} &\gate{H}& \qw & \qw & \qw & \ctrl{1} & \qw & \ctrl{1} &  \qw &\qw &\qw& \qw&\gate{H}&\qw\\
        & \wire[l][1]["\s_n"{above,pos=0.2}]{a} & \gate{H}&\qw & \qw & \qw &\targ{} & \gate{R_z(2t)} & \targ{} &  \qw & \qw &\qw& \qw&\gate{H}&\qw\\
    \end{quantikz}
\end{center}
The encryption operation in the main text corresponds to $t=\frac{\pi}{4}$. Therefore, we can implement the encryption operation with $4n$ two-qubit gates and $2n+4$ single-qubit unitary operations.

Now, we explain a decomposition of the decryption operation in Eq.~\eqref{eq:decoding_op} into two-qubit gates. We first decompose the decryption operation as follows:
\begin{align}
    U_{\mathrm{dec}}^{(n)}=\alpha_0\widetilde{V}_{\s_1\n_1}^\dag V_3V_2V_1\widetilde{V}_{\s_1\n_1},
\end{align}
where $\widetilde{V}_{\s_1\s_2}$ is a two-qubit unitary operation relating the computational basis and the Bell basis:
\begin{align}
    \widetilde{V}_{\s_1\n_1}\ket{\phi_\mu}_{\s_1\n_1}=\ket{\mu_1}_{\s_1}\ket{\mu_2}_{\n_1}
\end{align}
with a binary representation $\mu=\mu_1\mu_2$, and 
\begin{align}
    V_\mu\coloneqq \frac{\alpha_\mu}{\alpha_0}\ket{\mu_1\mu_2}\bra{\mu_1\mu_2}_{\s_1\n_1}\otimes \left(\bigotimes_{j=2}^n\sigma_\mu^{(\n_j)\top}\right)+\left(\mathbb{I}_{\s_1\n_1}-\ket{\mu_1\mu_2}\bra{\mu_1\mu_2}_{\s_1\n_1}\right)\otimes \mathbb{I}_{\n_2\cdots \n_n}.
\end{align}
Since $V_\mu$ can be implemented by controlled-controlled unitary operation, we have the following quantum circuits:
\begin{center}
    \begin{quantikz}
        V_1
    \end{quantikz}=
    \begin{quantikz}
        & \wire[l][1]["\s_1"{above,pos=0.2}]{a}&\gate{X} &\ctrl{2} &\ctrl{2}& \ \ldots & \qw &\ctrl{3}& &\ctrl{4}&\gate{X} &  \\
        &\wire[l][1]["\n_1"{above,pos=0.2}]{a}& \qw &\control{} &\control{}& \ \ldots&\qw&\control{}&\qw& \control{} &\qw&\qw\\
        & \wire[l][1]["\n_2"{above,pos=0.2}]{a}&\qw& \gate{\frac{\alpha_1}{\alpha_0}\mathbb{I}}&\gate{X} & \ \ldots&\wire[l][1]["\ddots"{below,pos=0.1}]{a} & & & & &\\
        &\wire[l][1]["\n_{n-1}"{above,pos=0.2}]{a} & \qw & \qw &\qw  & \ \ldots&  &\gate{X}& & \qw &\qw & \\
        & \wire[l][1]["\n_n"{above,pos=0.2}]{a} &\qw & \qw && \ \ldots &  &  &  \qw &\gate{X}& & 
    \end{quantikz}
\end{center}
\begin{center}
    \begin{quantikz}
        V_2
    \end{quantikz}=
    \begin{quantikz}
        & \wire[l][1]["\s_1"{above,pos=0.2}]{a} & &\ctrl{2} &\ctrl{2}& \ \ldots& \qw &\ctrl{3}& \ctrl{4}&\qw &\\
        &\wire[l][1]["\n_1"{above,pos=0.2}]{a} & \gate{X} & \control{} &\control{} & \ \ldots&\qw&\control{}&\control{} &\gate{X} &\qw\\
        & \wire[l][1]["\n_2"{above,pos=0.2}]{a}& & \gate{\frac{\alpha_2}{\alpha_0}\mathbb{I}}&\gate{-Y}& \ \ldots &\wire[l][1]["\ddots"{below,pos=0.1}]{a} & & & & \\
        &\wire[l][1]["\n_{n-1}"{above,pos=0.2}]{a} & & \qw  & \qw& \ \ldots & &\gate{-Y}&  &\qw  &\\
        & \wire[l][1]["\n_n"{above,pos=0.2}]{a} & & & & \ \ldots &  &  &  \gate{-Y}& &
    \end{quantikz}
\end{center}
\begin{center}

    \begin{quantikz}
        V_3
    \end{quantikz}=
    \begin{quantikz}
        & \wire[l][1]["\s_1"{above,pos=0.2}]{a} &\ctrl{2} &\ctrl{2}& \ \ldots& \qw &\ctrl{3}& &\ctrl{4}&\qw \\
        &\wire[l][1]["\n_1"{above,pos=0.2}]{a}  &\control{} &\control{}& \ \ldots&\qw&\control{}&\qw& \control{} &\\
        & \wire[l][1]["\n_2"{above,pos=0.2}]{a}&\gate{\frac{\alpha_3}{\alpha_0}\mathbb{I}}&\gate{Z}& \ \ldots &\wire[l][1]["\ddots"{below,pos=0.1}]{a} & & & & \\
        &\wire[l][1]["\n_{n-1}"{above,pos=0.2}]{a} & \qw & \qw&  \ \ldots & &\gate{Z}& & \qw &\qw \\
        & \wire[l][1]["\n_n"{above,pos=0.2}]{a} & \qw & & \ \ldots  & &  &  \qw &\gate{Z}& \\
    \end{quantikz}
\end{center}
Note that each controlled-controlled unitary operation can be implemented by five two-qubit gates \cite{sleator_realizable_1995,nielsen_quantum_2010,yu_five_2013}. Therefore, we can implement the decryption operation with $15n+7$ two-qubit gates. 

In summary, the whole protocol can be implemented with at most $21n+11$ two-qubit gates in total.

\end{document}